\documentclass[lettersize,journal]{IEEEtran}
\usepackage{amsmath,amssymb,amsfonts,bm} 
\usepackage{algorithm} 
\usepackage{array}
\usepackage{caption}
\usepackage{subcaption}
\usepackage{textcomp}
\usepackage{stfloats}
\usepackage{bigints}
\usepackage{booktabs}
\usepackage{url}
\usepackage{algpseudocode}
\usepackage{verbatim}
\usepackage{graphicx}
\algrenewcommand\algorithmicrequire{\textbf{Input:}}
\algrenewcommand\algorithmicensure{\textbf{Output:}}
\usepackage{mathtools}
\usepackage{dsfont}
\usepackage{esvect}
\usepackage{accents}
\usepackage{algpseudocode}
\usepackage{cite}
\usepackage{balance}
\usepackage{xcolor}
\newcolumntype{M}[1]{>{\centering\arraybackslash}m{#1}}
\newcommand*{\QEDA}{\null\nobreak\hfill\ensuremath{\blacksquare}}%

\newcommand\myeqa{\mathrel{\stackrel{\makebox[0pt]{\mbox{\normalfont\scriptsize (a)}}}{=}}}
\newcommand\myeqb{\mathrel{\stackrel{\makebox[0pt]{\mbox{\normalfont\scriptsize (b)}}}{=}}}
\newcommand\myeqc{\mathrel{\stackrel{\makebox[0pt]{\mbox{\normalfont\scriptsize (c)}}}{=}}}

\begin{document}

\title{Analysis of Integrated Communication and Sensing in Cylindrical Aerial
Corridors}
%
%
%

\author{Harris K.~Armeniakos,~\IEEEmembership{Member,~IEEE,} Petros S. Bithas,~\IEEEmembership{Senior Member,~IEEE,}\\
        Athanasios G.~Kanatas,~\IEEEmembership{Senior Member,~IEEE,} 
        and Harpreet S. Dhillon,~\IEEEmembership{Fellow,~IEEE}

\thanks{This work was co-funded by the University of Piraeus Research Center (UPRC) and the EU HORIZON programme under iSEE-6G GA No. 101139291} 
 \thanks{A preliminary version of this work was presented in \cite{GC26}.}
\thanks{The work of H. S. Dhillon was supported by US NSF Grants CNS-1923807 and CNS-2107276.}
\thanks{H. K. Armeniakos and A. G. Kanatas are with the Department
of Digital Systems, University of Piraeus, 18534 Piraeus, Greece (e-mail:
\{harmen,kanatas\}@unipi.gr).}
\thanks{P. S. Bithas is with the Department of Digital Industry
Technologies, National and Kapodistrian University of Athens,
34400 Psachna, Evia, Greece
(e-mail: pbithas@uoa.gr).}
\thanks{H. S. Dhillon is with Wireless@Virginia Tech, Bradley Department of
Electrical and Computer Engineering, Virginia Tech, Blacksburg, 24061, VA, USA (e-mail: hdhillon@vt.edu).}
}
\maketitle

\begin{abstract}
In unmanned aerial vehicle (UAV) networks, integrated communication and sensing (ISAC) is emerging as a key enabler to support extended and continuous communication and sensing capabilities for sixth-generation (6G) services among UAVs operating in swarms. In this paper, a comprehensive framework is presented to investigate the ISAC coverage probability  of a reference UAV-base station (BS) in a UAV network. A fixed number of UAV user equipment (UAV-UEs) and UAV-BSs are deployed in an aerial corridor modeled as a finite cylinder. The three-dimensional (3D) binomial point process (BPP) is used to model both the 3D spatial locations of the UAV-UEs and the UAV-BSs. By adopting realistic 3D third generation partnership project (3GPP) antenna patterns for both the UAV-UEs and UAV-BSs, the UAVs' boresight direction of the maximum antenna gain is modeled as a random unit vector. All UAV-BSs have sensing capabilities, so an arbitrary yet fixed reference UAV-BS  acts as a monostatic radar to acquire sensing measurements associated with a target  UAV-UE at a given time snapshot while the target  UAV-UE communicates with the reference UAV-BS. Consequently, considering both clutter and communication interference, exact analytical expressions for ISAC coverage probability are derived. Low complexity baseline analytical expressions are also derived by ignoring clutter interference and directional characteristics of the UAV-UEs. Finally, motivated by the vision for intelligent UAV networks, a DeepSets-based model is developed and trained to estimate the optimal location of the reference UAV-BS that maximizes ISAC performance given a spatial deployment of the UAVs. To the best of our knowledge, this is the first work to rigorously investigate ISAC performance in UAV networks while incorporating the directionality characteristics of the UAV antennas. 
\end{abstract}


%
\IEEEpeerreviewmaketitle

\section{Introduction}
As 6G networks evolve towards highly dynamic 3D architectures, UAV communications are expected to support aerial relaying, data delivery, energy transfer, and cooperative swarm operation \cite{alqudsi2025uav,swarms}. In this context, UAV-to-UAV (U2U) links are essential for distributed coordination, collision avoidance, relaying, and real-time information exchange without continuous dependence on terrestrial infrastructure \cite{Prabhath2025Spectrum3D}. Since the performance of such networks depends on the positions, mobility, and spatial concentration of the UAVs, accurate spatial modeling is necessary to analyze connectivity, interference, coverage, and scalability. Several spatial models have been proposed for UAV-assisted networks. Early finite-network models considered UAV-BSs uniformly distributed over bounded circular or 3D regions according to a binomial point process (BPP), allowing the characterization of serving/interference distances and coverage probability \cite{Chetlur2016FiniteUAV,Chetlur2017Finite3DUAV}. Cluster-process-based models were later adopted for UAV-enabled disaster-recovery networks, capturing the spatial concentration of aerial platforms and users around affected areas \cite{Hayajneh2018DisasterRecovery}. 3D deployment models based on truncated-octahedron tessellation were proposed to provide space-filling architectures for aerial BSs \cite{Mozaffari2019Beyond5G}, while spherical service-region models were used for UAV-to-everything networks with users randomly distributed around a serving UAV \cite{Hou2020U2XNOMA}. The truncated-octahedron framework was further generalized in \cite{Armeniakos2022TruncatedOctahedron} through finite BPP and finite homogeneous Poisson point process models for aerial nodes within each 3D cell.

In recent times, the increasing number of UAVs and traffic has necessitated the development of aerial highways or UAV corridors. These corridors restrict UAV movement to predefined one-dimensional (1D) or quasi-one-dimensional routes to prevent collisions and comply with regulatory requirements \cite{cherif20213d,karimi2024optimizing}. This allows connectivity resources to be concentrated along specific flight paths instead of providing uniform coverage throughout airspace. The cellular coverage of UAVs that move along corridors was analyzed in \cite{Maeng2024UAVCorridor}, while antenna tilt, transmit power, and infrastructure placement were optimized in \cite{KarimiBidhendi2024UAVCorridors}. In \cite{Armeniakos2024JointCoverage}, UAV-BSs were modeled by a 1D BPP over a finite corridor, and the joint energy and SINR coverage of RF-powered IoT devices was derived. Furthermore, \cite{Prabhath2025Spectrum3D} studied the spectral efficiency and spectrum utilization in 3D multi-UAV networks. Although aerial-corridor-based network design is receiving increasing attention \cite{10974465,10711847}, it remains at an early stage, with open issues related to corridor design, infrastructure optimization, traffic regulation, and performance evaluation.

\subsection{Related Work and Motivation}
UAV-enabled integrated sensing and communication (ISAC) has emerged as a promising framework to jointly support aerial connectivity, environmental sensing, target detection, and low-altitude traffic management. The main applications, technical challenges, and future directions of UAV-assisted ISAC were discussed in \cite{Mu2023UAVMeetsISAC,Meng2024UAVEnabledISAC}, highlighting the role of UAV mobility, flexible deployment, swarm operation, trajectory optimization, and joint resource allocation. In \cite{Hu2022TrajectoryRadar}, a cellular-connected UAV was used for communication-assisted radar sensing with trajectory optimization under sensing-resolution constraints, while \cite{Tropkina2023Distributed} developed a distributed communication and bistatic sensing architecture to improve collision avoidance and resilience. Stochastic-geometry-based ISAC models were considered in \cite{Maeng2023UAVRadarCoexistence,Zhang2024PerceptiveUAV}, where UAV radar/communication nodes, BSs, users, and aerial targets were modeled through spatial point processes to analyze ranging, communication, and surveillance performance. More recently, networked ISAC frameworks have been proposed for coordinated beamforming, UAV trajectory optimization, aerial-object sensing, continuous UAV tracking, and communication handover \cite{Cheng2025NetworkedISAC,Zhao2025TrackingHandover}. Although some of these works capture the random spatial distribution of UAVs, communication nodes, sensing targets, and interference sources, the behavior of the directional antenna of aerial nodes remains an equally critical aspect for aerial-link characterization.

In UAV-assisted networks, the desired and interfering powers depend on the angular position of the nodes, the beam-steering directions, the sidelobes, and the beam misalignment. Practical 3GPP or array-based antenna patterns have been incorporated into UAV trajectory optimization, clustered millimeter-wave UAV networks, wireless-backhauled UAV systems, SINR meta-distribution analysis, and UAV-corridor coverage studies \cite{Chowdhury2020Trajectory,Yi2020ClusteredUAV,Banagar2022TwoHop,Shi2023MetaDistribution,Maeng2024UAVCorridor,KarimiBidhendi2024UAVCorridors}. More recently, coordinated transmit beamforming, UAV trajectories, and BS associations were jointly optimized for networked ISAC with aerial communication and 3D sensing requirements \cite{Cheng2025NetworkedISAC}. These works show that realistic antenna and beamforming models are essential for capturing off-boresight attenuation, sidelobe interference, beam alignment, and the coupling between UAV position and link quality.

Stochastic geometry provides a powerful mathematical framework for modeling the spatial aspects of UAV networks and analyzing their performance. Although the Poisson point process (PPP) has widely been adopted mainly for the sake of mathematical tractability, it is not directly applicable  for modeling finite networks. When a fixed and predetermined number of UAVs are deployed in a finite region, their spatial distribution can be well described by the homogeneous BPP. Compared to a homogeneous PPP, BPP is neither ergodic nor stationary, and the BS's performance is location-dependent. For example, the aggregate interference and the SIR of a BS located in the center of the network differ significantly from a BS at the network edge. In addition, the distances of the UAVs to an arbitrary located UAV-BS are correlated because of the mutual distance between the UAV-BS and the center of the network. Because of these reasons, BPP allows for capturing the network's performance at an arbitrary location of the UAV-BS in the network. \emph{To the best of the authors’ knowledge, this is the first work to develop a stochastic geometry framework for the rigorous analysis of ISAC performance in aerial networks, with UAV-UE locations modeled as a 3D BPP in aerial cylindrical corridors. } 

\subsection{Contributions}
The main contributions of this paper are the following: 

\subsubsection{Novel 3D Spatial Modeling of UAV-UEs and UAV-BSs} A new stochastic geometry framework is developed to model the spatial distribution of UAVs in finite aerial networks. Balancing tractability and realism, the UAV-UEs and UAV-BSs are modeled as independent three-dimensional binomial point processes (BPPs) within a cylindrical UAV corridor. A reference UAV-BS is then placed at a fixed but otherwise arbitrary location within the corridor to characterize position-dependent performance in these networks.  

\subsubsection{ISAC Performance Analysis} 
The system performance is analyzed in terms of the ISAC coverage probability. Both UAV-UEs and UAV-BSs are modeled using realistic 3GPP-based antenna patterns, with the boresight direction corresponding to the maximum antenna gain of each UAV-UE represented as a random unit vector on the unit sphere. By adopting a shared spatial geometry for the sensing and communication functionalities, exact-form expressions for the ISAC coverage probability are derived. To provide further analytical tractability, simplified expressions are also obtained for a reduced scenario in which backscatter interference and the directional characteristics of the UAV-UE antennas are neglected.

\subsubsection{DeepSets-based Model for Optimizing the Location of reference UAV-BS} Having investigated the ISAC performance of an arbitrary reference UAV-BS, a direct and meaningful extension is to estimate the 3D position of the reference UAV-BS which maximizes the ISAC performance given the spatial geometry of the UAVs in the network. To enable intelligent UAV deployment, a DeepSets-based model is developed to map each spatial realization of UAVs in the corridor to the corresponding optimal location of the reference UAV-BS.

\subsubsection{System-level Insights \& Design Guidelines} Key system-level insights are derived by quantifying the impact of
backscatter interference and antenna-directionality characteristics on ISAC
performance. In addition, important design parameters of the UAV corridor are identified to maximize the ISAC coverage probability. Finally, the effect of the reference UAV-BS location on the ISAC coverage probability is
demonstrated, and its optimal placement is investigated.  

The remainder of the paper is organized as follows. In Section II, the system model and the antenna pattern modeling are presented. The SIR metric is also defined.  In Section III, the boresight modeling of the UAVs' maximum antenna gain is presented. Subsequently, the ISAC performance analysis is conducted. In Section IV, the simplified scenario is presented and the corresponding expressions of the ISAC coverage probability are derived. Section V presents the numerical results and the insights. Section VI presents the DeepSets-based model and the corresponding evaluation results, while Section VII concludes the paper.

\begin{figure*}[!t]
\centering
  \includegraphics[width=0.8\textwidth]{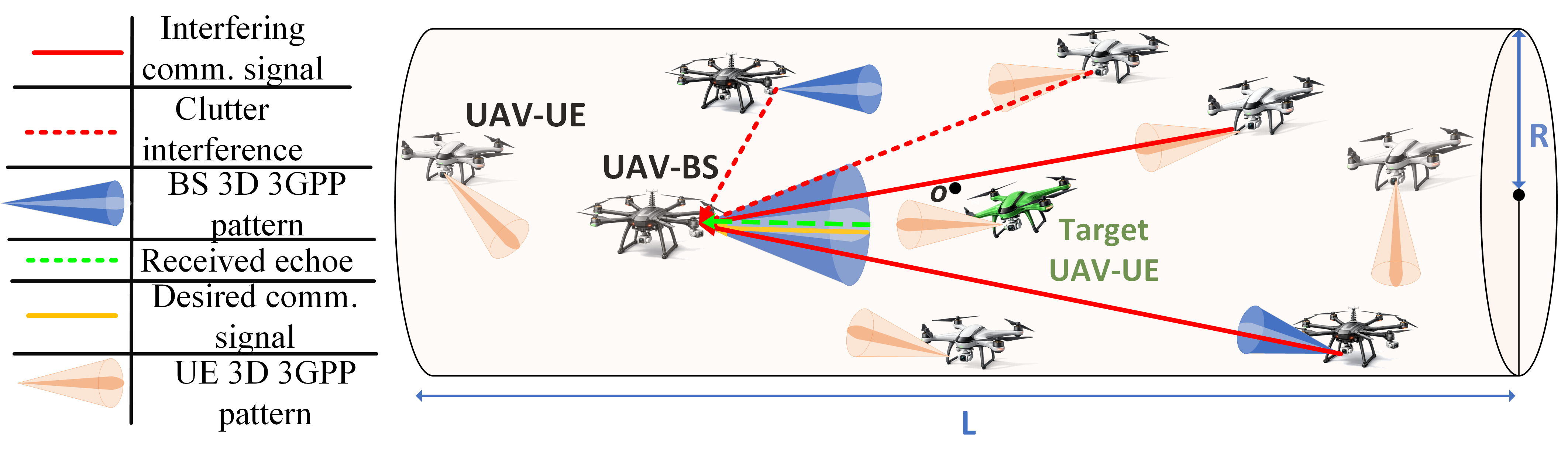}
  \caption{Illustration of the UAV ISAC-enabled network with a UAV-BS equipped with a 3D 3GPP-based antenna pattern and UAV-UEs forming a 3D BPP in a UAV corridor.}
  \label{fig:concept_corridor}
\end{figure*}

\section{System Model}
\subsection{Spatial Modeling of Network Nodes}
Consider an ISAC-enabled UAV network consisting of  multiple UAV-UEs and  UAV-BSs. Assume that both UAV-UEs and UAV-BSs hover in an UAV aerial corridor with rules enforced by appropriate traffic authorities, as will be discussed next. The UAV corridor is modeled as a finite \emph{3D cylinder}, which has been proposed as one of the most reasonable geometric models to approximate a real UAV corridor \cite{yu2025hybrid}. Consequently, it is assumed that both the UAV-UEs and the UAV-BSs are independently and uniformly distributed within the finite cylindrical region $\mathcal{C}(L,R)$, where $\mathcal{C}(L,R)$ represents a 3D cylinder centered at the origin $\mathbf{o}$, with axial extent from $-L/2$ to $L/2$ and radius $R$. In particular, $\mathcal{C}(L,R) = \{(x,y,z)\in \mathbb{R}^3 : -\frac{L}{2} \leq x\leq\frac{L}{2}, y^2+z^2 \leq R^2\}$ with volume $|\mathcal{C}(L,R)| = \pi R^2 L$. Therefore, a predefined and fixed number of $N_{UE}$ UAV-UEs is assumed to be distributed on $\mathcal{C}(L,R)$ and their spatial locations form a 3D BPP $\Phi_{UE} = \{\mathbf{p}_1,...,\mathbf{p}_{N_{UE}}\}$ with $\{\mathbf{p}_i\}=\{[x_i^{UE},y_i^{UE},z_i^{UE}]^T\}_{i=1:N_{UE}}$. Similarly, a fixed number of $N_{BS}$ UAV-BSs is assumed to be distributed on $\mathcal{C}(L,R)$ and their spatial locations form a 3D BPP $\Phi_{BS}=\{\mathbf{q}_1,...,\mathbf{q}_{N_{BS}}\}$  with $\{\mathbf{q}_j\}=\{[x_j^{BS},y_j^{BS},z_j^{BS}]^T\}_{i=1:N_{BS}}$. 
In addition to the $N_{BS}$ UAV-BSs, one additional UAV-BS is designated as controller, responsible for coordinating the necessary inter-UAV signaling and enforcing compliance with the relevant air-traffic regulations. This UAV-BS will henceforth be referred to as the \emph{reference UAV-BS} with index $\{0\}$ and it is assumed to be deployed in a known and fixed 3D location $\mathbf{q}_0 = [x_0^{BS},y_0^{BS},z_0^{BS}]^T$. Let $ \mathbf{r}_i^{\mathrm{UE}} = \mathbf{p}_i - \mathbf{q}_0$  define the vector from the reference UAV-BS to the $i$-th UAV-UE and $d_i^{\mathrm{UE}}=\|\mathbf{r}_i^{\mathrm{UE}}\|$ define its Euclidean distance. Similarly, let $\mathbf{r}_j^{\mathrm{BS}} = \mathbf{q}_j - \mathbf{q}_0$ define the vector from the reference UAV-BS to the $j$-th UAV-BS and $d_j^{\mathrm{BS}}=\|\mathbf{r}_j^{\mathrm{BS}}\|$ define its Euclidean distance.


\subsection{Antenna Modeling}
All UAVs (UAV-UEs and UAV-BSs) in the network exploit beamforming techniques to enable ISAC capabilities. The UAV antenna array can produce a directive beam both for the sensing and communication functionalities. In this work, a realistic 3GPP-based antenna pattern is adopted, and thus the actual antenna gain within the 3dB beamwidth range is not constant. When directed toward an angle pair $(\theta_0,\phi_0)$, this antenna pattern has loss in elevation,  loss in azimuth as well as a  3D radiation power patterns (in dB) given respectively by
\begin{align}\label{eqpatterns}
       &L_{3gpp,k}^e(\theta,\theta_0) = {\rm{min}}\Bigg\{ 12\Bigg(\frac{\theta-\theta_0}{\theta_{3dB,k}} \Bigg)^2, SLA \Bigg\}, \\ \nonumber
       &L_{3gpp,k}^a(\phi,\phi_0) = {\rm{min}}\Bigg\{ 12\Bigg(\frac{\phi-\phi_0}{\phi_{3dB,k}} \Bigg)^2, SLA \Bigg\},  \\ \nonumber
       &_{3gpp,k}^{3D}(\theta,\theta_0,\phi,\phi_0) =\\ \nonumber
       &G_{{\rm{max}}}-{\rm{min}}\Big\{L_{3gpp,k}^e(\theta,\theta_0) + L_{3gpp,k}^a(\phi,\phi_0),SLA\Big\} \nonumber,     
 \end{align}
where $k \in \{\mathrm{BS},\mathrm{UE}\}$, $G_{{\rm{max}}}$ denotes the maximum antenna gain and $\theta_{3dB,k}$,  $\phi_{3dB,k}$ denote the elevation and azimuthal $3$~dB beamwidths of the UAV-UE and UAV-BS antennas and $SLA = 30$~dB is the front-to-back ratio. The multiplicative antenna gain factor is indicated by $g_{3gpp,k}^{3D}(\theta,\theta_0,\phi,\phi_0)$ and the multiplicative maximum antenna gain factor is indicated by $g_{{\rm{max}}}$. For mathematical convenience, let $g_{3gpp,k}^{3D}(\Delta \theta,\Delta \phi)$ denote the multiplicative antenna gain factor of a UAV-UE or a UAV-BS w.r.t. elevation and azimuthal angular distances $\Delta \theta = \theta-\theta_0$ and $\Delta \phi =  \phi-\phi_0$, respectively, as mentioned in \eqref{eqpatterns}. 

Sensing measurements associated with a target UAV-UE are used to estimate the relative angular direction of the UAV-UE with respect to the reference UAV-BS, allowing the beam to steer toward the target UAV-UE \cite{10806828}. This reflects the practical ISAC operation, in which beam alignment is performed using angular estimation. Thus, the target UAV-UE continuously experiences the maximum UAV-BS antenna gain, while all other UAVs are seen through off-boresight gains acting as clutter interference during the sensing phase.

\subsection{Path Loss and Channel Modeling}
A key characteristic of UAV networks is the ability of UAVs to establish line-of-sight (LoS) links above rooftop level. In this context, following the standard power-law path-loss model for the path between the $i$-th UAV-UE and the reference UAV-BS, the  path-loss function  is denoted by $l(\|\mathbf{r}_i^{\mathrm{UE}}\|) = {d_{\mathrm{UE},i}}^{-\alpha}$, with $\alpha$ denoting the path loss exponent for the LoS propagation links. Similarly, the path-loss function for the link between the $j$-th UAV-BS and the reference UAV-BS is given by $l(\|\mathbf{r}_j^{\mathrm{BS}}\|) = {d_{\mathrm{BS},j}}^{-\alpha}$. 

The amplitudes of the communication and sensing channels are assumed to experience Nakagami-$m$ fading. Accordingly, the power gains $h^{o}_{\mathrm{UE},i}$, $o\in\{c,s\}$, associated with the links between the $i$-th UAV-UE and the reference UAV-BS follow the Gamma distribution. 
For the links between the $j$-th UAV-BS and the reference UAV-BS, $h^{e}_{\mathrm{BS},j}$ and $h^{s}_{\mathrm{BS},j}$ denote the independent channel power gains associated with the reflected-echo interference and the direct RF interference, respectively. Therefore,
$h^{o}_{\mathrm{UE},i},\,
h^{e}_{\mathrm{BS},j},\,
h^{s}_{\mathrm{BS},j}
\sim \mathrm{Gamma}\left(m,\frac{1}{m}\right)$,
with shape parameter $m$ and scale parameter $1/m$.

\subsection{ISAC Operation, Association Policy, and SIR Definition}
\subsubsection{ISAC Operation and UAV-UE Association Policy}
In this work, we focus on the ISAC performance of the reference UAV-BS which is assumed to act as a monostatic radar to acquire sensing measurements associated with a target  UAV-UE at a given time snapshot. During the sensing phase, the \emph{target UAV-UE} is chosen uniformly at random among the set of $N_{\mathrm{UE}}$ available UAV-UEs of the spatial realization $\Phi_{\mathrm{UE}}$,  generated in the considered time snapshot. Meanwhile, the other UAV-BSs perform sensing operations toward other UAV-UEs within the corridor.  Subsequently, in the communication phase, the target UAV-UE performs uplink communication  with the reference UAV-BS. Similarly, the remaining UAV-UEs also perform uplink communication with other UAV-BSs. Let $\mathbf{p}_0 = [x_0^{\mathrm{UE}},y_0^{\mathrm{UE}},z_0^{\mathrm{UE}}]^T$ denote the spatial location of the target UAV-UE. Once the position $\mathbf{p}_0 $ of the target UAV-UE is determined, let $\Phi_{\mathrm{UE}}^! = \Phi_{\mathrm{UE}} \setminus \{\mathbf{p}_0\}$. 

For the adopted scheduling mechanism of the network, it is assumed that sensing and communication functionalities are performed using different time-frequency resources. The reference UAV-BS acts as a monostatic radar  to acquire a sensing snapshot of the target UAV-UE, while the target UAV-UE performs reverse communication with the reference UAV-BS, as shown in Fig. 1.

\subsubsection{Communication and Sensing SIR Definition}
\paragraph{Definition of Sensing SIR} During the sensing phase, the reference UAV-BS measures the reflected echo from the target UAV-UE. In practice, the quality of the received sensing echo  determines how accurately the reference UAV-BS can estimate  radar sensing parameters, such as the target UAV-UE range and angular direction \cite{10938034}. At the same time, the reference UAV-BS suffers from i) aggregate clutter interference caused by the rest $(N_{\mathrm{UE}}-1)$ of the  UAV-UEs and the $N_{\mathrm{BS}}$ UAV-BSs in the corridor and ii) aggregate RF interference caused by the $N_{BS}$ UAV-BSs that perform the tracking operation. By following the radar equation, the  sensing SIR from the target UAV-UE received at the reference UAV-BS's antenna can be written as 
\begin{equation}\label{eqSIRs}
{\rm{SIR}}_s =  \frac{P_{\mathrm{echo}}}{I_s^{\rm{tot}}}=\frac{P_{\rm{echo}}}{I_s^{\mathrm{UE\, echo}} +I_s^{\mathrm{BS\, echo}} +I_s^{\mathrm{BS\, RF}}}, 
\end{equation}
where 
\begin{align} \label{sensingInt}
&P_{\mathrm{echo}} = p_t g_{{\rm{max}}}^2 h_{\mathrm{UE},0}^s \sigma_0 l(\|\mathbf{q}_0-\mathbf{p}_0\|)^2\frac{\lambda^2}{(4 \pi)^3}, \\ \nonumber
&I_s^{\mathrm{UE\, echo}} = \sum\limits_{i \in \Phi_{\mathrm{UE}}^!} p_t {(g_{i}^{\mathrm{0 \rightarrow UE}})}^2 h_{\mathrm{UE},i}^s \sigma_{\mathrm{UE},i}  l(\|\mathbf{r}_i^{\mathrm{UE}}\|)^2\frac{\lambda^2}{(4 \pi)^3},\\\nonumber
&I_s^{\mathrm{BS\, echo}} = \sum\limits_{j \in \Phi_{\mathrm{BS}}} p_t {(g_{j}^{\mathrm{0 \rightarrow BS}})}^2 h_{\mathrm{BS},j}^e \sigma_{\mathrm{BS},j}  l(\|\mathbf{r}_j^{\mathrm{BS}}\|)^2\frac{\lambda^2}{(4 \pi)^3},\\\nonumber
&I_s^{\mathrm{BS\, RF}} = \sum\limits_{j \in \Phi_{\mathrm{BS}}} p_t g_{j}^{\mathrm{0 \rightarrow BS}} g_{j}^{\mathrm{BS \rightarrow 0}} h_{\mathrm{BS},j}^s K_f l(\|\mathbf{r}_j^{\mathrm{BS}}\|).
\end{align}

In \eqref{sensingInt}, $p_t$ denotes the equal transmit power for all UAVs, $\lambda$ denotes the carrier wavelength, $\sigma_0$ denotes the random radar cross-section (RCS) area of the target UAV-UE, $\sigma_{\mathrm{BS},j}, \sigma_{\mathrm{UE},i}$ denote the RCS area of the $j$-th UAV-BS and the  $i$-th UAV-UE, respectively, and $K_f = (\lambda/4 \pi)^2$. By adopting the Swerling I model, the probability density function (PDF) of the RCS $\sigma_0,\sigma_{\mathrm{BS},j}, \sigma_{\mathrm{UE},i}$ is given by $f_{\sigma}(\sigma) = \frac{1}{\overline{\sigma}}\exp(-\sigma/\overline{\sigma})$, where $\overline{\sigma}$ denotes the average RCS.  Moreover, $h_{\mathrm{BS},j}^e, h_{\mathrm{BS},j}^s$ in the interference terms $I_s^{\mathrm{BS\, echo}},I_s^{\mathrm{BS\, RF}}$, respectively, denote the power gains of the channels between the $j$-th UAV-BS and the reference UAV-BS.

For the antenna gains in \eqref{sensingInt}, we define the gain of the reference UAV-BS toward the $i$-th interfering UAV-UE, the gain of the reference UAV-BS toward the $j$-th interfering UAV-BS and the gain of the $j$-th interfering UAV-BS toward the reference UAV-BS, respectively, as 
\begin{align} \label{gains}
&g_{i}^{\mathrm{0 \rightarrow UE}} = g_{3gpp,BS}^{3D}(\Delta \theta_i^{\mathrm{0 \rightarrow UE}},\Delta \phi_i^{\mathrm{0 \rightarrow UE}}),\\ \nonumber
&g_{j}^{\mathrm{0 \rightarrow BS}} = g_{3gpp,BS}^{3D}(\Delta \theta_j^{\mathrm{0 \to BS}},\Delta \phi_j^{\mathrm{0 \rightarrow BS}}),\\ \nonumber
&g_{j}^{\mathrm{BS \rightarrow 0}} = g_{3gpp,BS}^{3D}(\Delta \theta_j^{\mathrm{BS \to 0}},\Delta \phi_j^{\mathrm{BS \rightarrow 0}}), 
\end{align}
with corresponding elevation and azimuthal angular distances given by $\Delta \theta_i^{\mathrm{0 \rightarrow UE}}, \Delta \theta_i^{\mathrm{0 \rightarrow BS}},\Delta \theta_j^{\mathrm{BS \rightarrow 0}}$ and $\Delta \phi_i^{\mathrm{0 \rightarrow UE}}, \Delta \phi_j^{\mathrm{0 \rightarrow BS}}, \Delta \phi_j^{\mathrm{BS \rightarrow 0}}$, respectively.

\paragraph{Definition of Communication SIR} During the communication phase, it is assumed that the target UAV-UE is performing reverse communication with the reference UAV-BS, while the reference UAV-BS suffers from RF interference caused by the rest $(N_{\mathrm{UE}}-1)$ active UAV-UEs.  The rest of the UAV-BSs are assumed to be in reception mode during this phase. All the remaining $(N_{\mathrm{UE}}-1)$ UAV-UEs are assumed to be active, which represents a worst-case uplink interference scenario.\footnote{The proposed framework is directly applicable to scenarios in which only a subset of the $(N_{\mathrm{UE}}-1)$ UAV-UEs is assumed to be active. This is captured by assuming that a portion of the UAV-UEs is in sleep mode.} 
The communication SIR from the target UAV-UE received at the reference UAV-BS's antenna array is given by
\begin{equation}\label{eqSIRc}
{\rm{SIR}}_c = \frac{S_0}{I_c^{\mathrm{UE\, RF}}},
\end{equation}
where
\begin{align}\label{Ic}
    &S_0 =g_{{\rm{max}}}^2 h_{\mathrm{UE},0}^c l(\|\mathbf{q}_0-\mathbf{p}_0\|),\\ \nonumber
    &I_c^{\mathrm{UE\, RF}}=\sum\limits_{i \in \Phi_{\mathrm{UE}}^{!}} g_{i}^{\mathrm{0 \rightarrow UE}} g_{i}^{\mathrm{UE \rightarrow 0}} h_{\mathrm{UE},i}^c l(\|\mathbf{r}_i^{\mathrm{UE}}\|), \\ \nonumber
    &g_{i}^{\mathrm{UE \rightarrow 0}} = g_{3gpp,UE}^{3D}(\Delta \theta_i^{\mathrm{UE \rightarrow 0}},\Delta \phi_i^{\mathrm{UE \rightarrow 0}}),\\
\end{align}
denote the signal power received from the target UAV-UE, the aggregate interference power and the gain of the $i$-th interfering UAV-UE toward the reference UAV-BS with elevation and azimuthal angular distances given by $\Delta \theta_i^{\mathrm{UE \rightarrow 0}},\Delta \phi_i^{\mathrm{UE \rightarrow 0}}$, respectively.

\section{ISAC Performance Analysis}
In this section, the ISAC performance analysis is conducted in terms of the ISAC coverage probability.  

\emph{Definition 1:} \emph{The ISAC CP $\mathcal{P}_{cs}(\theta_c, \theta_s)$, is defined as the probability that the received SIR of the communication and sensing functions at the receiving UAV-BS are simultaneously above a predefined threshold $\theta_c$ and $\theta_s$ required for successful communication and tracking, respectively, i.e., $\mathcal{P}_{cs}(\theta_c, \theta_s)  \triangleq \mathbb{P}[ {\rm{SIR}}_{c}>\theta_c, {\rm{SIR}}_s>\theta_s]$. }

Notably, the sensing and communication functions share the same underlying spatial realization of the UAV-UEs and UAV-BSs. Consequently, \(\mathrm{SIR}_c\) and \(\mathrm{SIR}_s\) are statistically dependent random variables. Therefore, the ISAC coverage probability must be expressed as the joint probability $ \mathcal P_{cs}(\theta_c,\theta_s)=
\mathbb P\left[{\rm SIR}_c>\theta_c,\ {\rm SIR}_s>\theta_s\right]$ and, in general, cannot be decomposed into the product of the marginal coverage probabilities.

\subsection{Modeling and Statistical Characterization of Antenna Boresight Direction for UAV-BSs and UAV-UEs}
Before proceeding with the derivation of the ISAC coverage probability, a statistical model for the UAV-BSs and UAV-UEs antenna boresight direction is first introduced. Subsequently, the angular distances involved in \eqref{gains} and \eqref{Ic} are statistically characterized. 

Each UAV-BS is assumed to steer its beam toward a target UAV-UE during the
sensing phase, while the corresponding UAV-UE performs reverse communication
under beam alignment with its associated UAV-BS. Hence, for each realization
of $\Phi_{\rm UE}$ and $\Phi_{\rm BS}$, the boresight directions of the
UAV-BSs and UAV-UEs are determined by their own target associations and
scheduling decisions, and are generally not aligned with the reference
UAV-BS. From the viewpoint of the reference UAV-BS, these directions are
unknown and random; therefore, following the random-interfering-beam modeling approach in \cite{rebato2019stochastic},
they are modeled as random unit vectors on the unit sphere. This isotropic
orientation model provides a tractable random-beam approximation for
capturing off-boresight interference. Let $\mathbb{S}^2 \triangleq \{\mathbf{u} \in \mathbb{R}^3: \|\mathbf{u}\|=1\}$ define the unit sphere. For the $i$-th UAV-UE and the $j$-th UAV-BS, the boresight vector is respectively given by 
\begin{equation}
\mathbf{b}_i^{\mathrm{UE}}=
\begin{bmatrix}
b_{x,i}^{\mathrm{UE}} \\
b_{y,i}^{\mathrm{UE}}  \\
b_{z,i}^{\mathrm{UE}}
\end{bmatrix}, \,\,\, 
\mathbf{b}_j^{\mathrm{BS}}=
\begin{bmatrix}
b_{x,j}^{\mathrm{BS}} \\
b_{y,j}^{\mathrm{BS}}  \\
b_{z,j}^{\mathrm{BS}}
\end{bmatrix}.
\end{equation}
Accordingly, $\mathbf{b}_i^{\mathrm{UE}}, \mathbf{b}_j^{\mathrm{BS}} \sim \mathrm{Unif}(\mathbb{S}^2)$. Then, the azimuth and elevation angles of the antenna boresight direction of the $i$-th UAV-UE and the $j$-th UAV-BS are, respectively, given by
\begin{align} \label{anglesb}
    &\phi_{b,i}^{\mathrm{UE}} = \mathrm{atan2}(b_{y,i}^{\mathrm{UE}},b_{x,i}^{\mathrm{UE}}), \\ \nonumber
    &\theta_{b,i}^{\mathrm{UE}} = \mathrm{atan2}\Big(b_{z,i}^{\mathrm{UE}},\sqrt{(b_{x,i}^{\mathrm{UE}})^2 + (b_{y,i}^{\mathrm{UE}})^2}\Big), \\ \nonumber
    &\phi_{b,j}^{\mathrm{BS}} = \mathrm{atan2}(b_{y,j}^{\mathrm{BS}},b_{x,j}^{\mathrm{BS}}), \\ \nonumber
    &\theta_{b,j}^{\mathrm{BS}} = \mathrm{atan2}\Big(b_{z,j}^{\mathrm{BS}},\sqrt{(b_{x,j}^{\mathrm{BS}})^2 + (b_{y,j}^{\mathrm{BS}})^2}\Big),
\end{align}
where $\mathrm{atan2}(\cdot)$ denotes the four-quadrant arctangent function \cite{press2007numerical}. Next, let the unit vector denoting the direction from $j$-th UAV-BS toward the reference UAV-BS and the direction from $i$-th UAV-UE toward the reference UAV-BS be, respectively, defined as 
\begin{align}
   &\mathbf{v}_j^{\mathrm{BS}} = -\frac{\mathbf{r}_j^{\mathrm{BS}}}{\|\mathbf{r}_j^{\mathrm{BS}}\|} = [v_{x,j}^{\   \mathrm{BS}},v_{y,j}^{\mathrm{BS}},v_{z,j}^{\mathrm{BS}}]^T, \\ \nonumber
   &\mathbf{v}_i^{\mathrm{UE}} = -\frac{\mathbf{r}_i^{\mathrm{UE}}}{\|\mathbf{r}_i^{\mathrm{UE}}\|} = [v_{x,i}^{\mathrm{UE}},v_{y,i}^{\mathrm{UE}},v_{z,i}^{\mathrm{UE}}]^T.
\end{align}
Then, the azimuth and elevation angles of $\mathbf{v}_j^{\mathrm{BS}},\mathbf{v}_i^{\mathrm{UE}}$, are respectively, given by 
\begin{align} \label{anglesnodes}
    &\phi_{v,j}^{\mathrm{BS}} = \mathrm{atan2}(v_{y,j}^{\mathrm{BS}},v_{x,j}^{\mathrm{BS}}),\\ \nonumber
    &\theta_{v,j}^{\mathrm{BS}} = \mathrm{atan2}\Big(v_{z,j}^{\mathrm{BS}},\sqrt{(v_{x,j}^{\mathrm{BS}})^2+(v_{y,j}^{\mathrm{BS}})^2}\Big),\\ \nonumber
     &\phi_{v,i}^{\mathrm{UE}} = \mathrm{atan2}(v_{y,i}^{\mathrm{UE}},v_{x,i}^{\mathrm{UE}}),\\ \nonumber
    &\theta_{v,i}^{\mathrm{UE}} = \mathrm{atan2}\Big(v_{z,i}^{\mathrm{UE}},\sqrt{(v_{x,i}^{\mathrm{UE}})^2+(v_{y,i}^{\mathrm{UE}})^2}\Big). 
\end{align}
Then, by exploiting \eqref{anglesb} and \eqref{anglesnodes}, we define the azimuthal and elevation angular distances $\Delta \phi_j^{\mathrm{BS \rightarrow 0}}, \Delta \theta_j^{\mathrm{BS \rightarrow 0}}$ and $\Delta \phi_i^{\mathrm{UE \rightarrow 0}}, \Delta \theta_i^{\mathrm{UE \rightarrow 0}}$ as 
\begin{align} \label{deltangles}
    &\Delta \phi_j^{\mathrm{BS \rightarrow 0}} = [\phi_{v,j}^{\mathrm{BS}}-\phi_{b,j}^{\mathrm{BS}}]_{\pi},\\ \nonumber
    &\Delta \theta_j^{\mathrm{BS \rightarrow 0}} = \theta_{v,j}^{\mathrm{BS}}-\theta_{b,j}^{\mathrm{BS}},\\ \nonumber
     &\Delta \phi_i^{\mathrm{UE \rightarrow 0}} = [\phi_{v,i}^{\mathrm{UE}}-\phi_{b,i}^{\mathrm{UE}}]_{\pi},\\ \nonumber
    &\Delta \theta_i^{\mathrm{UE \rightarrow 0}} = \theta_{v,i}^{\mathrm{UE}}-\theta_{b,i}^{\mathrm{UE}},
\end{align}
where $[x]_{\pi}=\mathrm{mod}(x + \pi, 2\pi) - \pi$. In particular, given the antenna boresight directions of the UAV-BSs and UAV-UEs, the angular distances defined in \eqref{deltangles} determine the UAV-BSs' and UAV-UEs' antenna gain toward the reference UAV-BS, involved in \eqref{gains} and \eqref{Ic}, respectively. Having defined the angular distances in \eqref{deltangles}, the conditional joint PDF of the angular distances $\Delta \phi_j^{\mathrm{BS \rightarrow 0}},\Delta \theta_j^{\mathrm{BS \rightarrow 0}}$ or $\Delta \phi_i^{\mathrm{UE \rightarrow 0}},\Delta \theta_i^{\mathrm{UE \rightarrow 0}}$ is derived as an intermediate step in the ISAC coverage probability analysis.

\emph{Lemma 1:} \emph{Conditioned on $\Phi_k$, let $\mathbf{s}_{n}^{k}=[s_{x,n}^{k},s_{y,n}^{k},s_{z,n}^{k}]$ denote the spatial location of an interfering UAV of type $k\in\{\mathrm{UE},\mathrm{BS}\}$, where $n=i$ for UAV-UEs and $n=j$ for UAV-BSs. Then, the conditional joint PDF of the angular distances $\Delta\phi_{n}^{k\to0}$ and $\Delta\theta_{n}^{k\to0}$ is}
\begin{equation}\label{joint_pdf_offsets}
f_{\Delta\phi_{n}^{k\to0},\Delta\theta_{n}^{k\to0}\mid\Phi_k}(\phi,\theta)
=
\frac{1}{4\pi}
\cos\left(
\epsilon_0(\mathbf{s}_{n}^{k})-\theta
\right),
\end{equation}
\emph{for $\phi\in[-\pi,\pi],  \theta\in [\epsilon_0(\mathbf{s}_{n}^{k})-\frac{\pi}2,\epsilon_0(\mathbf{s}_{n}^{k})+\frac{\pi}2]$, where  $\epsilon_0(\mathbf{s}_{n}^{k})=\mathrm{atan2}\left(z_0^{\mathrm{BS}}-s_{z,n}^{k}, \sqrt{ (x_0^{\mathrm{BS}}-s_{x,n}^{k})^2+ (y_0^{\mathrm{BS}}-s_{y,n}^{k})^2 } \right)$}.

\textit{Proof.} Refer to Appendix A. \QEDA

\subsection{Sensing Coverage Probability}
The derivation process of the ISAC coverage probability presented in this subsection is particularly challenging and departs from the similar derivations presented in prior works in two major aspects: i) Due to the use of 3D antenna  patterns, the statistical characterization of the ISAC coverage probability depends not only on the Euclidean distances among the UAVs, but also on their relative angular coordinates, and ii) the sensing and communication functionalities share the same spatial model, i.e., identical locations of UAV-UEs and UAV-BSs. Consequently, from a mathematical point of view, the derivation of the marginal sensing and communication coverage probability requires a careful  treatment.  

 Given the fixed location $\mathbf{q}_0$ of the reference UAV-BS, the conditional Laplace transform of the aggregate interference power experienced during the detection phase can now be derived.

\emph{Lemma 2:} \emph{Conditioned on $\Omega = \{\mathbf{p}_0,\Phi_{\mathrm{UE}}^!,\Phi_{\mathrm{BS}}\}$, the conditional Laplace transform of $I_s^{\rm{tot}}$ is given by}
\begin{align} \label{LTI}
    &\mathcal{L}_{I_s^{\rm{tot}}}(s|\Omega) = \prod_{\mathbf{p}_i \in \Phi_{\mathrm{UE}}^!} \Psi_m \left( \frac{s P_i^{\mathrm{UE}}(\mathbf{p}_0,\mathbf{p}_i) \overline{\sigma}}{m}\right) \\
    &\times  \prod_{\mathbf{q}_j \in \Phi_{\mathrm{BS}}} \mathcal{A}_j^{\mathrm{BS}}(s|\mathbf{p}_0,\mathbf{q}_j) \prod_{\mathbf{q}_j \in \Phi_{\mathrm{BS}}} \Psi_m \left( \frac{s P_j^{\mathrm{BS}}(\mathbf{p}_0,\mathbf{q}_j) \overline{\sigma}}{m}\right),
\end{align}
\emph{where  $P_i^{\mathrm{UE}}(\mathbf{p}_0,\mathbf{p}_i) = \frac{(g_{i}^{\mathrm{0 \rightarrow UE}})^2}{(d_i^{\mathrm{UE}})^{2 \alpha}} \frac{\lambda^2}{(4\pi)^3}$ with $g_{i}^{\mathrm{0 \rightarrow UE}}$ being given by \eqref{gains} where $\Delta\theta_i^{\mathrm{0 \to UE}}$, $\Delta\phi_i^{\mathrm{0 \to UE}}$, are given by \eqref{LT1}, \eqref{LT2}, shown at the bottom of the page, $\mathcal{A}_j^{\mathrm{BS}}(s|\mathbf{p}_0,\mathbf{q}_j)$ is given by \eqref{LT3}, shown at the bottom of the page, $P_i^{\mathrm{BS}}(\mathbf{p}_0,\mathbf{q}_j) = \frac{(g_{i}^{\mathrm{0 \rightarrow BS}})^2}{(d_j^{\mathrm{BS}})^{2\alpha}} \frac{\lambda^2}{(4\pi)^3}$ with $g_{j}^{\mathrm{0 \to BS}}$ being given by \eqref{gains}, where $\Delta\theta_j^{\mathrm{0 \to BS}}$, $\Delta\phi_j^{\mathrm{0 \to BS}}$ are given by \eqref{LT4}, \eqref{LT5}, shown at the bottom of the page,  and $\Psi_m(x)=\int_0^{\infty}e^{-u}(1+x u)^{-m}\mathrm{d}u$.}

\textit{Proof.} Please refer to Appendix B. \QEDA

\begin{figure*}[!b]
\hrulefill
\begin{equation} \label{LT1}
\Delta\theta_i^{\mathrm{0 \to UE}} = \mathrm{atan2}\left(z_i^{\mathrm{UE}}-z_0^{\mathrm{BS}}, \sqrt{(y_i^{\mathrm{UE}}-y_0^{\mathrm{BS}})^2+(x_i^{\mathrm{UE}}-x_0^{\mathrm{BS}})^2}\right)-\mathrm{atan2}\left(z_0^{\mathrm{UE}}-z_0^{\mathrm{BS}}, \sqrt{(y_0^{\mathrm{UE}}-y_0^{\mathrm{BS}})^2+(x_0^{\mathrm{UE}}-x_0^{\mathrm{BS}})^2}\right)
\end{equation}
\begin{equation}  \label{LT2}
 \Delta\phi_i^{\mathrm{0 \to UE}} = \left[\mathrm{atan2}\left(y_i^{\mathrm{UE}}-y_0^{\mathrm{BS}}, x_i^{\mathrm{UE}}-x_0^{\mathrm{BS}}\right)- \mathrm{atan2}\left(y_0^{\mathrm{UE}}-y_0^{\mathrm{BS}}, x_0^{\mathrm{UE}}-x_0^{\mathrm{BS}}\right) \right]_{\pi}
\end{equation}
\begin{equation}  \label{LT3}
    \mathcal{A}_j^{\mathrm{BS}}(s|\mathbf{p}_0,\mathbf{q}_j) =  \int_{\epsilon_0(\mathbf{q}_j)-\pi/2}^{\epsilon_0(\mathbf{q}_j)+\pi/2} \int_{-\pi}^{\pi}\left( 1+\frac{sg_{j}^{\mathrm{0 \rightarrow BS}} g_{j}^{\mathrm{BS \rightarrow 0}} K_f l(\|\mathbf{r}_j^{\mathrm{BS}}\|)}{m} \right)^{-m} f_{\Delta\phi_{j}^{\mathrm{BS}\to 0},\Delta\theta_{j}^{\mathrm{BS}\to0} \mid\Phi_{\mathrm{BS}}}(\Delta\phi,\Delta\theta) \mathrm{d} \Delta\phi  \mathrm{d} \Delta\theta
\end{equation}
\begin{equation} \label{LT4}
    \Delta\theta_j^{\mathrm{0 \to BS}} = \mathrm{atan2}\left(z_i^{\mathrm{BS}}-z_0^{\mathrm{BS}}, \sqrt{(y_i^{\mathrm{BS}}-y_0^{\mathrm{BS}})^2+(x_i^{\mathrm{BS}}-x_0^{\mathrm{BS}})^2}\right)-\mathrm{atan2}\left(z_0^{\mathrm{UE}}-z_0^{\mathrm{BS}}, \sqrt{(y_0^{\mathrm{UE}}-y_0^{\mathrm{BS}})^2+(x_0^{\mathrm{UE}}-x_0^{\mathrm{BS}})^2}\right)
\end{equation}
\begin{equation}  \label{LT5}
   \Delta\phi_j^{\mathrm{0 \to BS}} = \left[\mathrm{atan2}\left(y_j^{\mathrm{BS}}-y_0^{\mathrm{BS}}, x_j^{\mathrm{BS}}-x_0^{\mathrm{BS}}\right)- \mathrm{atan2}\left(y_0^{\mathrm{UE}}-y_0^{\mathrm{BS}}, x_0^{\mathrm{UE}}-x_0^{\mathrm{BS}}\right) \right]_{\pi} 
\end{equation}
\end{figure*}

Next, the conditional sensing coverage probability in terms of Lemma 2 can be derived.

\emph{Proposition 1:} \emph{Conditioned on the random set $\Omega = \{\mathbf{p}_0,\Phi_{\mathrm{UE}}^!,\Phi_{\mathrm{BS}}\}$, the conditional sensing coverage probability $\mathcal{P}_{s}(\theta_s|\Omega)$ of the reference UAV-BS is given by}
\begin{equation} 
\begin{split}
&\mathcal{P}_{s}(\theta_s|\Omega) \\
&= \int_{0}^{\infty} \sum_{k=0}^{m - 1}  \frac{(-s_0(\mathbf{p}_0))^k }{ k!} {\Bigg[ \frac{\partial^{k} \mathcal{L}_{I_s^{\rm{tot}}}(s|\Omega) }{\partial{s^k}}\Bigg]}_{s_0}f_{\sigma}(\sigma_0) {\rm{d}}  \sigma_0,  
\end{split}
\end{equation}
\emph{for $s_0(\mathbf{p}_0) = \frac{m \theta_s (4 \pi)^3}{\sigma_0 g_{{\rm{max}}}^2l^2(\|\mathbf{p}_0\|) \lambda^2}$}.

\textit{Proof.} The proof follows conceptual lines similar to those presented in \cite[Theorem 1]{TWC} and thus is omitted here for brevity.  \QEDA

\subsection{Communication Coverage Probability}
The conditional Laplace transform of the aggregate interference power distribution is  first derived.

\emph{Lemma 3:} \emph{Conditioned on $\mathbf{p}_0,\Phi_{\mathrm{UE}}^!$, the conditional Laplace transform  $\mathcal{L}_{I_c^{\mathrm{UE\, RF}}}(s|\mathbf{p}_0,\Phi_{\mathrm{UE}}^!)$ of the aggregate RF interference power distribution is given by}
\begin{equation}
     \mathcal{L}_{I_c^{\mathrm{UE\, RF}}}(s|\mathbf{p}_0,\Phi_{\mathrm{UE}}^!) = \prod_{\mathbf{p}_i \in \Phi_{\mathrm{UE}}^!} \mathcal{A}_i^{\mathrm{UE}}(s|\mathbf{p}_0,\mathbf{p}_i),
\end{equation}
\emph{where  
\begin{align}
     &\mathcal{A}_i^{\mathrm{UE}}(s|\mathbf{p}_0,\mathbf{p}_i)\\ \nonumber
     &= \int_{\epsilon_0(\mathbf{p}_i)-\pi/2}^{\epsilon_0(\mathbf{p}_i)+\pi/2} \int_{-\pi}^{\pi}\left( 1+\frac{sg_{i}^{\mathrm{0 \rightarrow UE}} g_{i}^{\mathrm{UE \rightarrow 0}} l(\|\mathbf{r}_i^{\mathrm{UE}}\|)}{m} \right)^{-m} \\ \nonumber
     & \times f_{\Delta\phi_{i}^{\mathrm{UE}\to 0},\Delta\theta_{i}^{\mathrm{UE}\to0}\mid\Phi_{\mathrm{UE}}}(\Delta\phi, \Delta\theta) \mathrm{d}\Delta\phi \mathrm{d}\Delta\theta.
\end{align}
}

\textit{Proof.} The proof  follows a similar procedure as the proof for deriving  $\mathcal{L}_{I_s^\mathrm{BS\, RF}}(s|\mathbf{p}_0,\Phi_{\mathrm{BS}})$ in Lemma 2. The only difference is  the averaging over $\Delta \phi_i^{\mathrm{UE \to 0}},\Delta \theta_i^{\mathrm{UE \to 0}}$ involved now in $g_{i}^{\mathrm{UE \rightarrow 0}}$ w.r.t. the joint PDF $ f_{\Delta\phi_{i}^{\mathrm{UE}\to 0},\Delta\theta_{i}^{\mathrm{UE}\to0}\mid\Phi_{\mathrm{UE}}}(\Delta\phi,\Delta\theta)$. \QEDA

 Now, the conditional communication coverage probability can be derived in terms of Lemma 3.

\emph{Proposition 2:} \emph{Conditioned on  $\mathbf{p}_0,\Phi_{\mathrm{UE}}^!$, the conditional communication coverage probability $\mathcal{P}_{c}(\theta_c|\mathbf{p}_0,\Phi_{\mathrm{UE}}^!)$ of the reference UAV-BS is given by}
\begin{align} 
&\mathcal{P}_{c}(\theta_c|\mathbf{p}_0,\Phi_{\mathrm{UE}}^!) \\ \nonumber
&= \sum_{k=0}^{m - 1}  \frac{(-s_0(\mathbf{p}_0))^k }{
k!} {\Bigg[ \frac{\partial^{k} \mathcal{L}_{I_c^{\mathrm{UE\, RF}}}(s|\mathbf{p}_0,\Phi_{\mathrm{UE}}^!)}{\partial{s^k}}\Bigg]}_{s_0}, \nonumber 
\end{align}
\emph{for $s_0(\mathbf{p}_0) = \frac{m \theta_c }{ g_{{\rm{max}}^2} l(\|\mathbf{q}_0-\mathbf{p}_0\|)}$}. 

\textit{Proof.} The proof  follows conceptual lines similar to those of the proof for deriving the conditional communication coverage probability presented in \cite[Proposition 3]{10711847}.\QEDA

\emph{Remark 1:} \emph{The results obtained in Propositions 1 and  2 can be exploited to directly investigate the marginal sensing and communication coverage probabilities, respectively, for a fixed or random 3D position of the target UAV-UE.} 

\subsection{ISAC Coverage Probability}
The ISAC coverage probability can now be derived in the following Theorem by exploiting Proposition 1 and Proposition 2. 

\emph{Theorem 1:} \emph{The ISAC coverage probability of a reference UAV-BS in a UAV network is given by}
\begin{align} 
 &\mathcal{P}_{cs}(\theta_c, \theta_s) =  f_{\mathbf{p}_0}(\mathbf{p}_0) (f_{\mathbf{p}_i}(\mathbf{p}_i))^{{N_{\mathrm{UE}}-1}} (f_{\mathbf{q}_j}(\mathbf{q}_j))^{N{\mathrm{BS}}}\\ \nonumber
 &\times \int_{\mathcal{C}}\mathrm{d} \mathbf{p}_0 \prod_{i=1}^{{N_{\mathrm{UE}}-1}} \int_{\mathcal{C}}\mathrm{d} \mathbf{p}_i \prod_{j=1}^{
 {N_{\mathrm{BS}}}} \int_{\mathcal{C}}\mathrm{d} \mathbf{q}_j \\ \nonumber
 &\times \mathcal{P}_{s}(\theta_s|\mathbf{p}_0,\Phi_{\mathrm{UE}}^!,\Phi_{\mathrm{BS}})\mathcal{P}_{c}(\theta_c|\mathbf{p}_0,\Phi_{\mathrm{UE}}^!),
\end{align}
\emph{where $f_{\mathbf{p}_0}(\mathbf{p}_0)=f_{\mathbf{p}_i}(\mathbf{p}_i)=f_{\mathbf{q}_j}(\mathbf{q}_j)=\frac{1}{|\mathcal{C}(L,R)|}$}.
 
\textit{Proof.} Refer to Appendix C. \QEDA

\emph{Remark 2:} \emph{The main source of analytical complexity in Theorem 1 is the spatial coupling between the sensing and communication functionalities introduced by the same underlying spatial process. If the spatial locations of UAV-UEs and UAV-BSs are assumed to change between sensing and communication, the computational complexity of Theorem 1 is dramatically reduced.} 

\section{ISAC Performance Analysis: Special Cases}
 Although Theorem 1 provided an exact expression of the ISAC coverage probability under a stochastic geometry framework that reasonably approximates real-world ISAC functionalities, its analytical complexity is high  due to the  spatial coupling between sensing and communication. To obtain further insights and provide useful benchmark results, this section investigates two practically relevant special cases. Specifically, we consider a simplified framework in which i) clutter interference is ignored and ii) isotropic radiation patterns are adopted for the UAV-UEs, with constant antenna gain equal to $g_{{\rm{max}}}$. These assumptions lead to more tractable expressions, enable the separate assessment of clutter interference and UAV-UE antenna directionality, and provide baseline results for comparison with the full ISAC framework. The first assumption is commonly adopted in the literature \cite{10938034,10048770}, whereas the second simplification is rational when UAV-UEs have limited beamforming capabilities.

\subsection{Communication and Sensing SIR}
Under the simplified framework, the reference UAV-BS suffers from aggregate RF interference caused by the $N_{BS}$ UAV-BSs that perform tracking operation. Accordingly, the  sensing SIR from the target UAV-UE received at the reference UAV-BS's antenna can be rewritten by recalling \eqref{eqSIRs} as 
\begin{equation}
{\rm{SIR}}_s =\frac{P_{\rm{echo}}}{I_s^{\mathrm{BS\, RF}}}. 
\end{equation}
 
Under the simplified framework, the communication SIR from the target UAV-UE received at the reference UAV-BS's antenna array is given by modifying \eqref{eqSIRc} as
\begin{equation}
{\rm{SIR}}_c = \frac{S_0}{I_c^{\mathrm{UE\, RF}}} = \frac{g_{{\rm{max}}} h_{\mathrm{UE},0}^c l(\|\mathbf{q}_0-\mathbf{p}_0\|)}{\sum\limits_{i \in \Phi_{\mathrm{UE}}^{!}} g_{i}^{\mathrm{0 \rightarrow UE}}  h_{\mathrm{UE},i}^c l(\|\mathbf{r}_i^{\mathrm{UE}}\|)}.
\end{equation}

\subsection{Sensing Coverage Probability}

The conditional Laplace transform of the aggregate interference power distribution experienced in the sensing phase under the simplified framework is first derived.

\emph{Corollary 1:}  \emph{Conditioned on $\mathbf{p}_0,\Phi_{\mathrm{BS}}$, the conditional Laplace transform of $I_s^{\rm{tot}}$ is given by}
\begin{align} \label{LTI}
    &\mathcal{L}_{I_s^{\mathrm{BS\, RF}}}(s|\mathbf{p}_0,\Phi_{\mathrm{BS}}) =  \prod_{\mathbf{q}_j \in \Phi_{\mathrm{BS}}} \mathcal{A}_j^{\mathrm{BS}}(s|\mathbf{p}_0,\mathbf{q}_j),
\end{align}
\emph{where $\mathcal{A}_j^{\mathrm{BS}}(s|\mathbf{p}_0,\mathbf{q}_j)$ is given by \eqref{LT3}.}

\textit{Proof.} The proof follows from the proof for deriving Lemma 2. \QEDA

\emph{Corollary 2:} \emph{Conditioned on $\mathbf{p}_0,\Phi_{\mathrm{BS}}$, the conditional sensing coverage probability $\mathcal{P}_{s}(\theta_s|\mathbf{p}_0,\Phi_{\mathrm{BS}})$ of the reference UAV-BS is given by}
\begin{equation} 
\begin{split}
&\mathcal{P}_{s}(\theta_s|\mathbf{p}_0,\Phi_{\mathrm{BS}}) \\
&= \int_{0}^{\infty} \sum_{k=0}^{m - 1}  \frac{(-s_0(\mathbf{p}_0))^k }{ k!} {\Bigg[ \frac{\partial^{k} \mathcal{L}_{I_s^{\mathrm{BS\, RF}}}(s|\mathbf{p}_0,\Phi_{\mathrm{BS}}) }{\partial{s^k}}\Bigg]}_{s_0}\\
&\times f_{\sigma}(\sigma_0) {\rm{d}}  \sigma_0,  
\end{split}
\end{equation}
\emph{for $s_0(\mathbf{p}_0) = \frac{m \theta_s (4 \pi)^3}{\sigma_0 g_{{\rm{max}}}^2 l^2(\|\mathbf{p}_0\|) \lambda^2}$}.

\subsection{Communication Coverage Probability}
The conditional Laplace transform of the aggregate interference power distribution is first derived in the simplified framework. 

\emph{Corollary 3:} \emph{Conditioned on $\mathbf{p}_0,\Phi_{\mathrm{UE}}^!$, the conditional Laplace transform  $\mathcal{L}_{I_c^{\mathrm{UE\, RF}}}(s|\mathbf{p}_0,\Phi_{\mathrm{UE}}^!)$ of the aggregate interference power distribution is given by}
\begin{equation}
\begin{split}
     &\mathcal{L}_{I_c^{\mathrm{UE\, RF}}}(s|\mathbf{p}_0,\Phi_{\mathrm{UE}}^!) \\
     &= \prod_{\mathbf{p}_i \in \Phi_{\mathrm{UE}}^!} \left( 1+\frac{sg_{{\rm{max}}} \,g_{i}^{\mathrm{0 \rightarrow UE}} \, l(\|\mathbf{r}_i^{\mathrm{UE}}\|)}{m} \right)^{-m}.
     \end{split}
\end{equation}

\textit{Proof.} The proof follows directly from the proof for deriving Lemma 3  by simply substituting $g_{i}^{\mathrm{UE \rightarrow 0}}=g_{{\rm{max}}}$. \QEDA

\emph{Corollary 4:} \emph{Conditioned on  $\mathbf{p}_0,\Phi_{\mathrm{UE}}^!$, the conditional communication coverage probability $\mathcal{P}_{c}(\theta_s|\mathbf{p}_0,\Phi_{\mathrm{UE}}^!)$ of the reference UAV-BS is given by}
\begin{align} 
&\mathcal{P}_{c}(\theta_c|\mathbf{p}_0,\Phi_{\mathrm{UE}}^!) \\ \nonumber
&= \sum_{k=0}^{m - 1}  \frac{(-s_0(\mathbf{p}_0))^k }{
k!} {\Bigg[ \frac{\partial^{k} \mathcal{L}_{I_c^{\mathrm{UE\, RF}}}(s|\mathbf{p}_0,\Phi_{\mathrm{UE}}^!)}{\partial{s^k}}\Bigg]}_{s_0}, \nonumber 
\end{align}
\emph{for $s_0(\mathbf{p}_0) = \frac{m \theta_c }{ g_{{\rm{max}}} l(\|\mathbf{q}_0-\mathbf{p}_0\|)}$}.

\subsection{ISAC Coverage Probability}

The ISAC coverage probability under the simplified framework is derived as follows. 

\emph{Proposition 3:} \emph{The ISAC coverage probability of a reference UAV-BS in a UAV network under isotropic antenna patterns for UAV-UEs and by neglecting clutter interference from echoes is given by}
\begin{align} 
 &\mathcal{P}_{cs}(\theta_c, \theta_s) = \left(\frac{1}{|\mathcal{C}(L,R)|}\right)^{N_{\mathrm{UE}}+N{\mathrm{BS}}}\int_{\mathcal{C}}\mathrm{d} \mathbf{p}_0 \\ \nonumber
 &\times \prod_{i=1}^{{N_{\mathrm{UE}}-1}} \int_{\mathcal{C}}\mathrm{d} \mathbf{p}_i \prod_{j=1}^{
 {N_{\mathrm{BS}}}} \int_{\mathcal{C}}\mathrm{d} \mathbf{q}_j  \mathcal{P}_{s}(\theta_s|\mathbf{p}_0,\Phi_{\mathrm{BS}})\mathcal{P}_{c}(\theta_c|\mathbf{p}_0,\Phi_{\mathrm{UE}}^!).
\end{align}

\textit{Proof.} The proof for deriving Proposition 3 follows the same conceptual lines as the proof for deriving Theorem 1. \QEDA

\section{Numerical Results and Discussion}
In this section, numerical results are presented to evaluate the ISAC performance achieved in the proposed UAV network. The accuracy of the analytical results is verified by comparing them with the simulated ones obtained using MC simulations. The following parameters have been used, unless stated otherwise: $L = 100$~m, $R=5$, $\alpha = 2.2$ \cite{9115898}, $m = 2$,  $N_{\mathrm{UE}} = 10$, $N_{\mathrm{BS}} = 5$,  $(\theta_c,\theta_s) = (-10,-10)$~dB, $(\phi_{3dB,\mathrm{UE}},\theta_{3dB,\mathrm{UE}}) = (60^o,60^o)$ \cite{8713514}, $(\phi_{3dB,\mathrm{BS}},\theta_{3dB,\mathrm{BS}}) = (45^o,45^o)$ \cite{9221328}, $G_{\rm{max}}=8$~dBi, $SLA=30$~dB and $\overline{\sigma} = 0.7 m^2$ \cite{electronics11162502},  while the reference UAV-BS is assumed to be located at the origin. 

\begin{figure}[!t]
    \centering
    \includegraphics[keepaspectratio,width=3.45in]{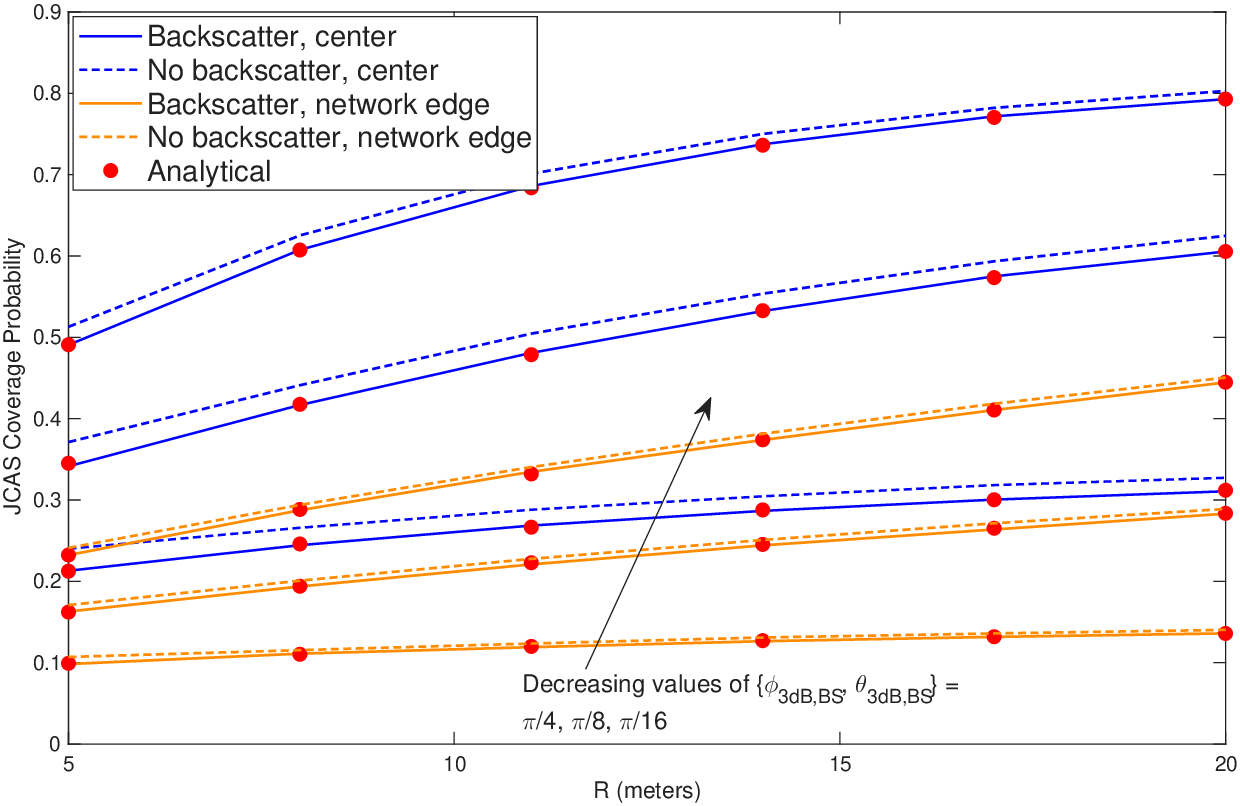}
       \caption{ISAC coverage probability versus $R$ for different values of $(\phi_{3dB,\mathrm{BS}},\theta_{3dB,\mathrm{BS}})$.}
    \label{fig:Fig2}
\end{figure}

Fig.~\ref{fig:Fig2} presents the analytical ISAC coverage probability
obtained from Theorem~1, denoted by markers, together with simulation
results, as a function of $R$ for different values of
$(\phi_{3dB,\mathrm{BS}},\theta_{3dB,\mathrm{BS}})$. To investigate the impact of backscatter interference on ISAC performance, the case in which
backscattering is ignored is also simulated. Two locations of the reference
UAV-BS are considered: i) at the center of the corridor and ii) at the edge
of the corridor, i.e., $[x_0^{BS},y_0^{BS},z_0^{BS}] = [L/2,0,0]$. The latter case is of
considerable practical interest, since edge-located UAV-BSs may arise in
corridor-entry or corridor-exit regions and typically experience asymmetric
interference conditions. A first observation is that, for both locations of the reference UAV-BS,
the ISAC coverage probability increases with the corridor radius $R$.
For a fixed antenna beamwidth, increasing the corridor radius $R$ increases
the angular separation among UAVs as observed from the reference UAV-BS,
thereby reducing the likelihood that interfering signals fall within its
main lobe. This increase becomes more evident for larger values of $R$ and is further enhanced as the UAV-BS antenna beamwidth decreases, since narrower beams
more effectively suppress off-boresight interference.
Interestingly, ignoring backscatter interference leads to a systematic overestimation of the ISAC performance. This trend is more pronounced for the case where the reference UAV-BS is located at the center of the corridor. The result also reveals that backscatter interference is highly location-dependent.

\begin{figure}[!t]
    \centering
    \includegraphics[keepaspectratio,width=3.45in]{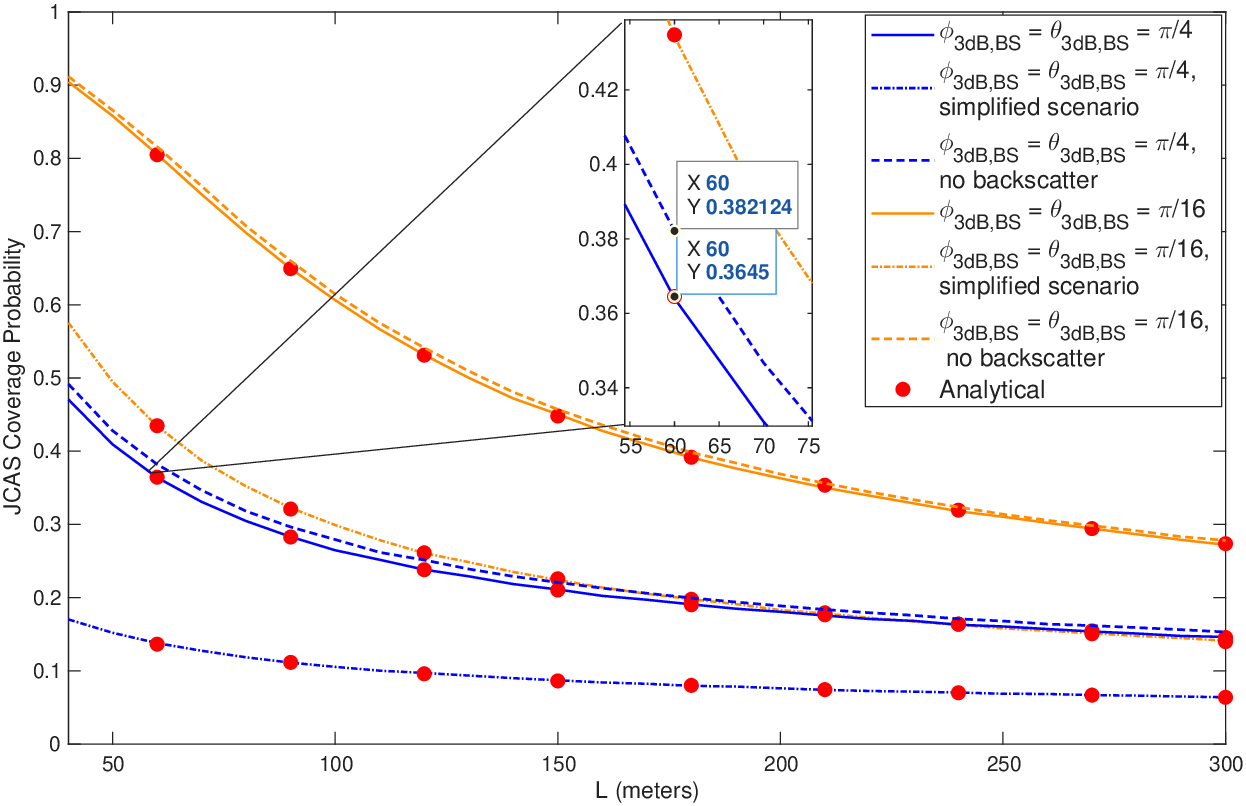}
       \caption{ISAC coverage probability versus $L$.}
    \label{fig:Fig3}
\end{figure}

\begin{figure}[!t]
    \centering
    \includegraphics[keepaspectratio,width=3.45in]{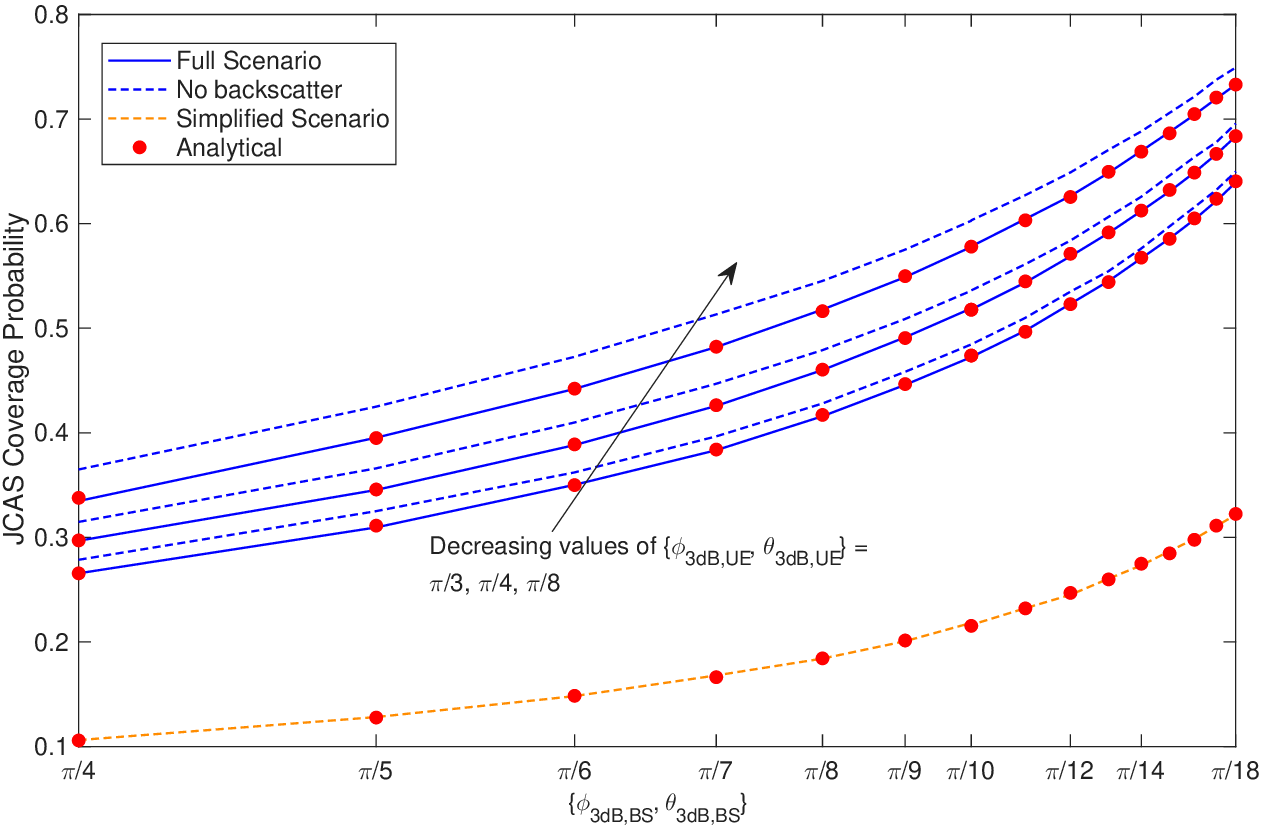}
       \caption{ISAC coverage probability versus $(\phi_{3dB,\mathrm{BS}},\theta_{3dB,\mathrm{BS}})$ for different values of $(\phi_{3dB,\mathrm{UE}},\theta_{3dB,\mathrm{UE}})$.}
    \label{fig:Fig4}
\end{figure}

\begin{figure}[!t]
    \centering
    \includegraphics[keepaspectratio,width=3.45in]{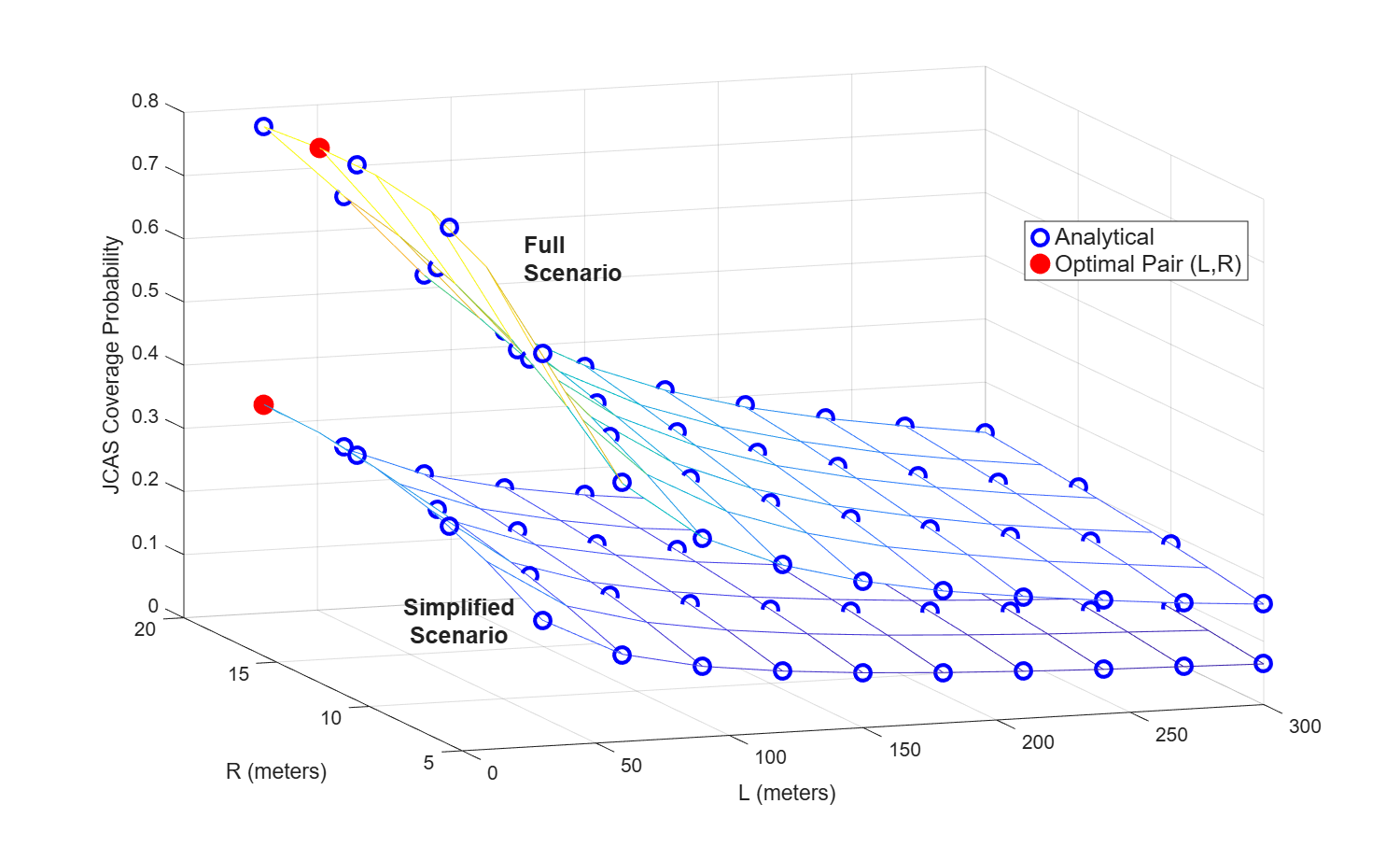}
       \caption{ISAC coverage probability versus $L$ and $R$.}
    \label{fig:Fig5}
\end{figure}

Fig. \ref{fig:Fig3} presents the  analytical (denoted by markers and obtained from Theorem 1 and Proposition 3) and simulated ISAC coverage probability versus $L$ for two values of $(\phi_{3dB,\mathrm{BS}},\theta_{3dB,\mathrm{BS}})$. Three scenarios are considered: i) The full scenario, ii) the scenario in which only the backscatter interference is ignored, and iii) the simplified scenario. A first observation is that ignoring backscatter interference leads to overestimating the ISAC performance. This is more pronounced for the wider beams of the UAV-BSs' antenna and for the smaller corridor lengths. Moreover, it is observed that, for wider UAV-BS antenna beams, the reduction
in the ISAC coverage probability with increasing $L$ rapidly saturates. For the simplified scenario, it is observed that the ISAC performance is considerably reduced. This behavior is mainly due to the assumption of non-directional UAV-UE
antennas, which significantly affects the ISAC coverage probability. Performance
 degradation is more pronounced for smaller values of $L$, since
the reduced corridor length limits spatial separation between aerial nodes
and increases the impact of interference. This observation highlights the
need to jointly consider directional antenna design not only at the UAV-BSs
but also at the UAV-UEs, particularly in compact aerial corridors.

Fig. \ref{fig:Fig4} presents the  analytical (denoted by markers and obtained from Theorem 1 and Proposition 3) and simulated ISAC coverage probability versus $\{\phi_{3dB,\mathrm{BS}},\theta_{3dB,\mathrm{BS}}\}$ for different values of $(\phi_{3dB,\mathrm{UE}},\theta_{3dB,\mathrm{UE}})$. For this figure, we have set $(\theta_c,\theta_s) = (10,-20)$~dB. Three scenarios are considered: i) The full scenario, ii) the scenario where only the backscatter interference are ignored, and iii) the simplified scenario. It is observed that the ISAC coverage probability increases as the beams of the UAV-UEs' antennas become more directive. Notably, the improvement is not linear. The gain from beam narrowing becomes more pronounced as the beams of the UAV-BSs' antenna become highly directive. Indeed, the curves increase moderately from $\pi/4$ to around $\pi/8$, and then increase more rapidly as the beamwidth approaches $\pi/18$. This indicates that highly directive beams of the UAV-BSs' antennas can provide a substantial performance enhancement, whereas improving the directionality of the UAV-UEs' antenna does not considerably enhance the ISAC performance. Interestingly, neglecting backscatter interference leads to an overestimation
of ISAC performance, and this overestimation becomes more pronounced for
narrower UAV-UE antenna beams. Finally, when isotropic antennas are adopted
at the UAV-UEs, the ISAC performance is significantly degraded. This degradation is more noticeable for narrower UAV-BS antenna beams, since the
benefits of highly directional UAV-BS transmissions cannot be fully exploited
when the UAV-UE side remains isotropic.

Fig. \ref{fig:Fig5} presents the  analytical, denoted by markers and obtained from Theorem 1 and Proposition 3, and simulated ISAC coverage probability versus $L$ and $R$ for the full scenario and the simplified one. The optimal pair value $(L,R)$ for which the ISAC coverage probability is optimized, is also depicted. The first observation is that the corridor length has a stronger effect on ISAC performance than the corridor radius. An interesting observation is that the directionality at the UAV-UEs' antennas affects the optimal pair $(L,R)$. Indeed, when isotropic antennas are adopted for the UAV-UEs, the ISAC performance is maximized at the smallest value of $L$ and the largest value of $R$. In contrast, in the original scenario, the ISAC performance is maximized for the smallest value considered of $L$ and for a moderate value of $R$. This result clearly demonstrates that the directional characteristics of the antennas of aerial users play a crucial role in the design of the corridor to maximize the ISAC performance.

\section{Optimization of the Reference UAV-BS Location: A DeepSets-based Model}
 In the previous sections, ISAC performance was investigated for an arbitrary but deterministic reference UAV-BS location within the UAV corridor. In this section, we optimize the 3D location of the reference UAV-BS to maximize the ISAC coverage probability for a given spatial realization of the UAV-UEs and UAV-BSs. The resulting placement problem is challenging because the ISAC coverage objective is a highly nonlinear and generally non-convex function of the reference UAV-BS coordinates through both distance-dependent propagation and angle-dependent antenna gains. Moreover, its gradient with respect to the UAV-BS position is not available in a tractable form for each network realization. Therefore, we employ a DeepSets-based architecture, which is naturally suited to learning permutation-invariant mappings from unordered spatial point sets to a near-optimal UAV-BS location. Specifically, the proposed architecture is trained to learn the mapping from each spatial network realization to a near-optimal reference UAV-BS location. Once trained, the model can infer a geometry-aware UAV-BS placement for a new spatial realization with low online computational complexity. Subsequently, online inference results are presented to evaluate the effectiveness and performance of the proposed DeepSets-based placement model.

\subsection{The proposed DeepSets-based Model}
 For the training process of the proposed model, exhaustive search is employed offline to generate supervised labels by selecting, for each training spatial realization, the candidate reference UAV-BS location
that maximizes the estimated ISAC coverage probability.\footnote{It is worth noting that the ISAC coverage probability can be evaluated using either Theorem 1 or Proposition 3 via Monte Carlo (MC) integration, with a computational time comparable to that of conventional MC simulation.} This offline label-generation step can be performed for representative setups for which exhaustive search is computationally feasible. Once trained, the DeepSets model provides online inference through a single forward pass, thereby avoiding repeated exhaustive-search evaluations for every new spatial realization. This is particularly useful for moderate and large UAV network deployments, where exhaustive search becomes increasingly
computationally demanding. The labels are subsequently used to train the DeepSets model to approximate the geometry-aware placement rule with significantly lower online computational complexity. The model applies a shared encoder $\psi(\cdot)$ to $\mathbf p_0$, $\Phi_{\mathrm{UE}}^!$ and $\Phi_{\mathrm{BS}}$. Subsequently, the model applies mean and max pooling. Mean pooling is used to capture the average spatial geometry and interference,  while max pooling is used to capture dominant or extreme spatial features and interference. The latter is particularly crucial for interference-limited networks. After encoding and pooling, the model performs the final mapping between the global feature vector $\mathbf h=[\mathbf h_{\rm tar}, \mathbf h_{\rm UE}, \mathbf h_{\rm BS}]$ and the predicted reference UAV-BS location $\rho(\cdot)$ as $\widehat{\mathbf q}_0 = \rho \left( \mathbf h \right)$. For a mini-batch $\mathcal{B}$, with $B$ training samples, the mini-batch loss function measures how close the DeepSets-predicted reference UAV-BS locations are to the exhaustive-search optimal locations and it is given by 
\begin{equation}
 \mathcal L(\boldsymbol{\Theta})  = \frac{1}{B}
    \sum_{b=1}^{B}
    \left\|
    \widehat{\mathbf q}_{0}^{(b)}
    -
    (\mathbf q_{0}^{\star})^{(b)}
    \right\|^2, 
\end{equation}
where $\widehat{\mathbf q}_{0}^{(b)}$ is the DeepSets prediction for the $b$-th sample in the mini-batch and $(\mathbf q_{0}^{\star})^{(b)}$ is the corresponding exhaustive-search label.

\begin{algorithm}[t]
\caption{DeepSets-Based Reference UAV-BS Placement}
\label{alg:deepsets_uavbs_placement}
\begin{algorithmic}[1]
\Require Number of training samples $D$; number of epochs $E$; mini-batch size $B$; candidate set $\mathcal Q\subset\mathcal C$; MC sample size $M$; thresholds $\theta_c,\theta_s$.
\Ensure Trained DeepSets placement model $f_{\boldsymbol{\Theta}}(\cdot)$.

\State Initialize the trainable parameters $\boldsymbol{\Theta}$ of $\psi_{\rm tar}(\cdot)$, $\psi_{\rm UE}(\cdot)$, $\psi_{\rm BS}(\cdot)$, and $\rho(\cdot)$.

\For{$d=1$ to $D$}
\State Generate one BPP realization of UAV-UEs:
\Statex \hspace{\algorithmicindent}
$\Phi_{\rm UE}^{(d)}={\mathbf p_0^{(d)},\mathbf p_1^{(d)},\ldots,\mathbf p_{N-1}^{(d)}}$.

\State Let $\mathbf p_0^{(d)}$ denote the target UAV-UE and define:
\Statex \hspace{\algorithmicindent}
$\Phi_{\rm UE}^{!,(d)}=\Phi_{\rm UE}^{(d)}\setminus\{\mathbf p_0^{(d)}\}$.

\State Generate one BPP realization of interfering UAV-BSs:
\Statex \hspace{\algorithmicindent}
$\Phi_{\rm BS}^{(d)}=\{\mathbf q_1^{(d)},\ldots,\mathbf q_{N_{\rm BS}}^{(d)}\}$.

\State Obtain the exhaustive-search label:
\Statex \hspace{\algorithmicindent}
$\begin{aligned}
&(\mathbf q_0^\star)^{(d)} \\
&=
\arg\max_{\mathbf q_0\in\mathcal Q}
\widehat{\mathcal P}_{cs} \left( \mathbf q_0;\theta_c,\theta_s
\right.\left.
\mid
\mathbf p_0^{(d)},(\Phi_{\rm UE}^{!})^{(d)},\Phi_{\rm BS}^{(d)}
\right).
\end{aligned}$

\State Store the training sample
$\left(\mathbf p_0^{(d)},(\Phi_{\rm UE}^{!})^{(d)},\Phi_{\rm BS}^{(d)},(\mathbf q_0^{\star})^{(d)}\right)$.
\EndFor

\For{$e=1$ to $E$}
\For{each mini-batch $\mathcal B$ of size $B$}
\For{each sample $\left(\mathbf p_0,\Phi_{\rm UE}^{!},\Phi_{\rm BS},\mathbf q_0^\star\right)\in\mathcal B$}

        \State Encode the target UAV-UE:
        \Statex \hspace{\algorithmicindent}
        $\mathbf h_{\rm tar}=\psi_{\rm tar}(\mathbf p_0)$.
        
        \State Encode and aggregate the interfering UAV-UE set:
        \Statex \hspace{\algorithmicindent}
        $\displaystyle
        \mathbf h_{\rm UE}
        =
        \left[
        \frac{1}{|\Phi_{\rm UE}^{!}|}
        \sum_{\mathbf p_i\in\Phi_{\rm UE}^{!}}
        \psi_{\rm UE}(\mathbf p_i),
        \;
        \max_{\mathbf p_i\in\Phi_{\rm UE}^{!}}
        \psi_{\rm UE}(\mathbf p_i)
        \right]$.
        
        \State Encode and aggregate the interfering UAV-BS set:
        \Statex \hspace{\algorithmicindent}
        $\displaystyle
        \mathbf h_{\rm BS}
        =
        \left[
        \frac{1}{|\Phi_{\rm BS}|}
        \sum_{\mathbf q_j\in\Phi_{\rm BS}}
        \psi_{\rm BS}(\mathbf q_j),
        \;
        \max_{\mathbf q_j\in\Phi_{\rm BS}}
        \psi_{\rm BS}(\mathbf q_j)
        \right]$.
        
        \State Predict the reference UAV-BS location:
        \Statex \hspace{\algorithmicindent}
        $\displaystyle
        \widehat{\mathbf q}_0
        =
        \rho
        \left(
        \mathbf h_{\rm tar},
        \mathbf h_{\rm UE},
        \mathbf h_{\rm BS}
        \right)$.
    \EndFor
    
    \State Update $\boldsymbol{\Theta}$ by minimizing the mini-batch loss:
    \Statex \hspace{\algorithmicindent}
    $\displaystyle
    \mathcal L(\boldsymbol{\Theta})
    =
    \frac{1}{B}
    \sum_{b=1}^{B}
    \left\|
    \widehat{\mathbf q}_{0}^{(b)}
    -
    (\mathbf q_{0}^{\star})^{(b)}
    \right\|^2$.
\EndFor

\EndFor

\State \Return $f_{\boldsymbol{\Theta}}(\cdot)$.
\end{algorithmic}
\end{algorithm}

\begin{table}[t]
\centering
\caption{Online computational time comparison between  exhaustive search and the DeepSets-based model.}
\label{tab:computational_time}
\footnotesize
\begin{tabular}{c c c}
\toprule
\textbf{Realization} 
& \textbf{Exhaustive Search Time [s]} 
& \textbf{DeepSets Time [s]} \\
\midrule
1  & 89.744 & 0.0082296 \\
2  & 91.798 & 0.0052867 \\
3  & 91.051 & 0.0113440 \\
4  & 93.317 & 0.0037482 \\
5  & 91.331 & 0.0069082 \\
6  & 92.826 & 0.0023138 \\
7  & 91.312 & 0.0016904 \\
8  & 92.906 & 0.0020057 \\
9  & 91.313 & 0.0017107 \\
10 & 92.854 & 0.0016158 \\
\midrule
\textbf{Mean} & \textbf{91.845} & \textbf{0.0044853} \\
\textbf{Std.} & \textbf{1.1133} & \textbf{0.0033785} \\
\bottomrule
\end{tabular}
\end{table}

The overall procedure for obtaining the trained DeepSets model is summarized in Algorithm 1.

\subsection{Evaluation of the DeepSets-based Model}
In this subsection, we evaluate the capability of our trained model in finding the optimal location for the reference UAVs-BS to maximize the ISAC performance given the spatial geometry of UAVs in the network. For the offline training process of the DeepSets-based models, we have considered 500 training samples. Then 100 evaluation samples were applied and 50 online test samples. The learning rate has been set to $10^{-3}$. Fig. 6a  presents the optimal location of the reference UAV-BS given a random spatial geometry of the network. In this model, 10 UAV-UEs have been assumed, whereas the 3dB antenna beamwidths of the UAV-BS have been set to $\pi/8$. In Fig. 6b, the trained model assumes a denser deployment of UAV-UEs with 20 UAV-UEs but more directive beams for the UAV-BSs' antennas, i.e., $\{\phi_{3dB,\mathrm{BS}},\theta_{3dB,\mathrm{BS}}\}=\pi/16$. For both models, we also estimate the optimal location of the reference UAV-BS through exhaustive search. The estimated optimal locations generated by the trained models in both figures reveal the effectiveness of the models in accurately estimating the optimal locations for the reference UAV-BS that maximize ISAC performance. Indeed, notice that even with a small training data set, the DeepSets-based models accurately estimate the location of the reference UAV-BS which is also validated from the exhaustive search algorithm. Due to the geometry of the corridor as well as the directional-dependent metrics, the optimal location of the reference UAV-BS for maximizing the ISAC performance given the spatial geometry of the UAVs, is not obvious. In Table \ref{tab:computational_time}, the online computational time between
the exhaustive search and the DeepSets-based model is compared for 10 realizations. It is clearly observed that  the exhaustive search algorithm can identify the optimal location of the reference UAV-BS but its computational time is orders of magnitude higher as compared to the DeepSets-based model. Therefore, the DeepSets-based model i) serves as a first yet reasonable step towards the development of learning models for maximizing ISAC performance in intelligent UAV networks and ii) requires substantially fewer computational resources than exhaustive-search-based optimization. These results highlight the potential of learning-based placement for computationally efficient optimization in future UAV networks.

\begin{figure*}[!t]
\centering
\begin{subfigure}[t]{0.50\textwidth}
\centering
\includegraphics[
width=\linewidth,
trim={80 65 80 55},
clip
]{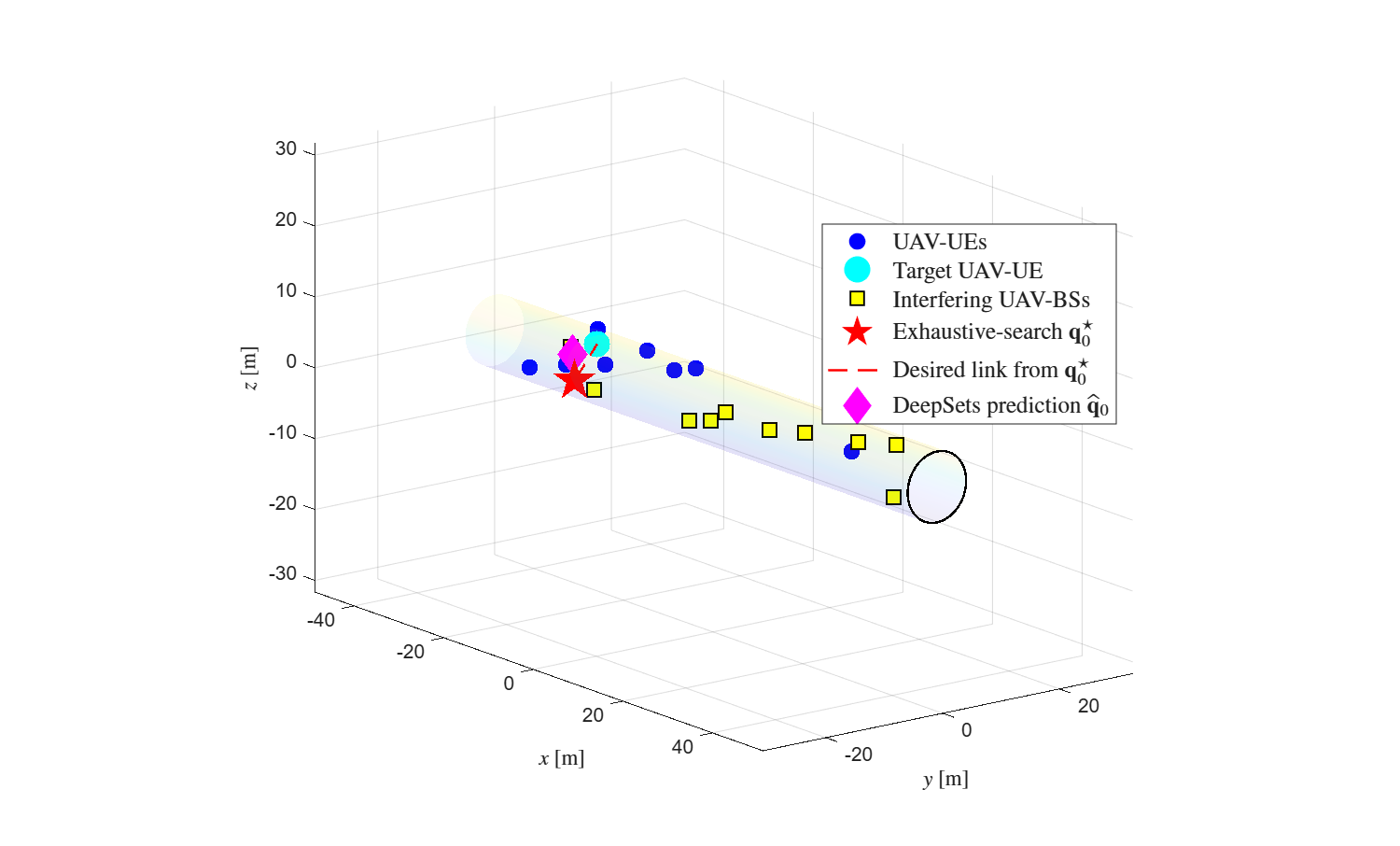}
\caption{10 UAV-UEs and $\{\phi_{3dB,\mathrm{BS}},\theta_{3dB,\mathrm{BS}}\} = \pi/8$.}
\label{fig:DL_pi8}
\end{subfigure}%
\hspace{-0.035\textwidth}%
\begin{subfigure}[t]{0.50\textwidth}
\centering
\includegraphics[
width=\linewidth,
trim={80 65 80 55},
clip
]{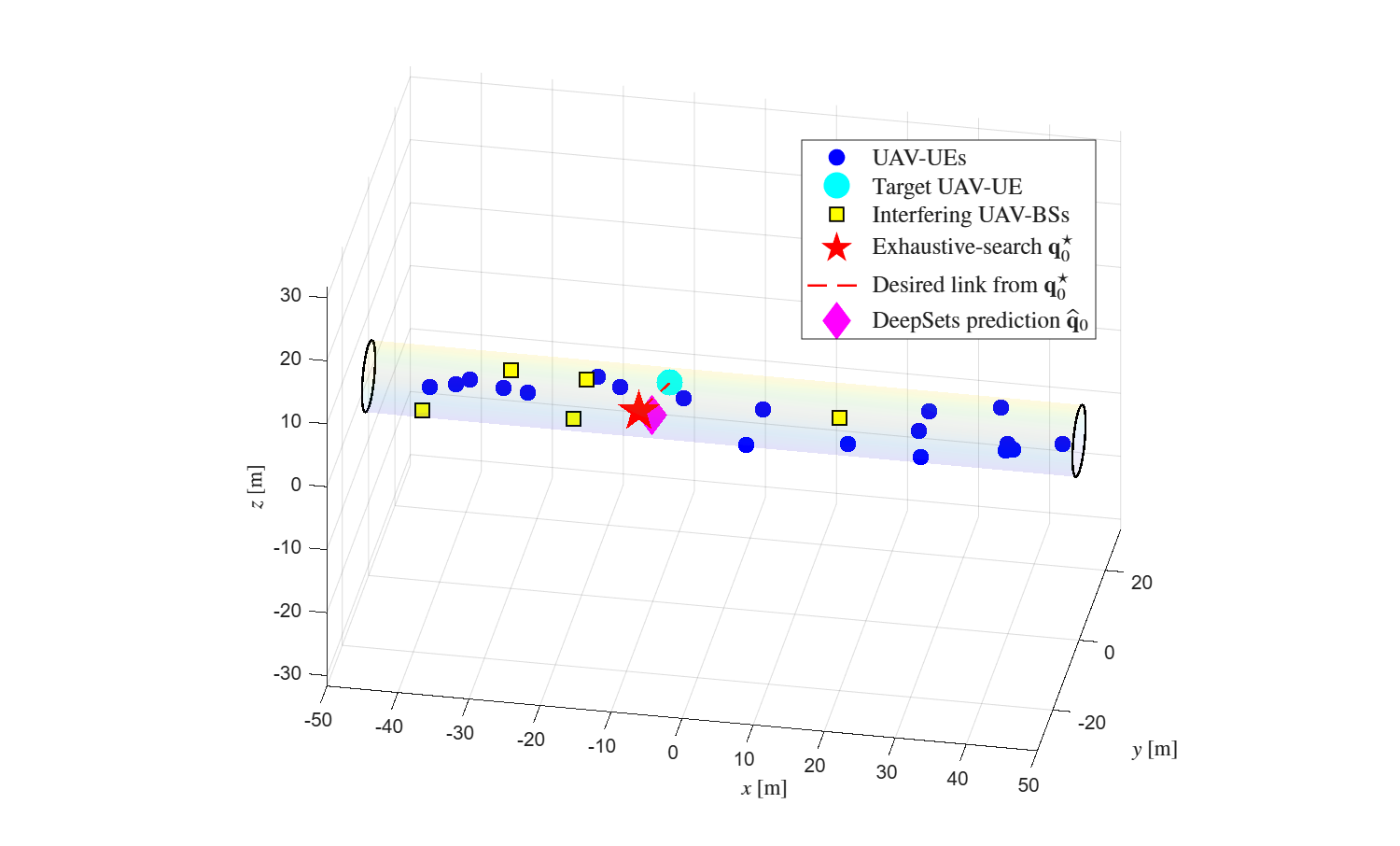}
\caption{20 UAV-UEs and $\{\phi_{3dB,\mathrm{BS}},\theta_{3dB,\mathrm{BS}}\} = \pi/16$.}
\label{fig:DL_pi16_20UAVs}
\end{subfigure}
\caption{Illustration of the optimal spatial deployment of the reference UAV-BS generated by the DeepSets-based model.}
\label{fig:DL_comparison}
\end{figure*}

\section{Conclusion}
A novel stochastic-geometry framework based on a 3D BPP was proposed to analyze the ISAC performance of
UAV corridor-enabled networks, where both UAV-UEs and UAV-BSs operate as
spatially distributed aerial swarms. Assuming that the UAVs are equipped with 3D 3GPP-based antenna patterns, the ISAC coverage probability was analytically characterized under clutter and RF interference. Analytical expressions for the simplified framework were also derived. Finally, a DeepSets-based model was developed and trained to estimate the reference UAV-BS's optimal location in the UAV corridor. Among several key system-level insights, it was shown that: i) ignoring backscatter interference causes overestimation of ISAC performance, especially for wider UAV-BS antenna beams and smaller corridors, ii) The ISAC coverage probability increases as the UAV-UEs’ antennas become more directional, but the gain from beam narrowing becomes more significant as the UAV-BSs’ antennas also become highly directional, and iii) the ISAC performance is maximized at the smallest value of $L$ and at a moderate value of $R$. Finally, a DeepSets-based model was trained to estimate the optimal location
of the reference UAV-BS, paving the way for the development of more
intelligent UAV swarm networks.

\appendices
\section{Proof of Lemma 1}
 Conditioned on $\Phi_k$, the spatial locations of all UAVs of type $k$ are known. Let $\mathbf{s}_{n}^{k}$ denote the fixed location of a UAV of type $k$. While the proof is  independent of the type of the UAV, we omit the type indexes $k,n$ for the rest of the proof and therefore $\mathbf{s}_{n}^{k}=\mathbf{s}$. We are interested in the statistical characterization of $\Delta\phi = [\phi(\mathbf{s})-\phi_b]_{\pi}$ and $\Delta\theta= \epsilon_0(\mathbf{s})-\theta_b$, where $\phi(\mathbf{s})=\mathrm{atan2}(y_0^{\mathrm{BS}}-s_y,x_0^{\mathrm{BS}}-s_x)$. The joint PDF of $f_{\phi_b,\theta_b}(\phi_b,\theta_b)$ is given by 
\begin{equation} \label{jointPDFangles}
    f_{\phi_b,\theta_b}(\phi_b,\theta_b) = \frac{1}{2 \pi} \frac{1}{2}\cos(\theta_b), 
\end{equation}
for $\phi_b \in [-\pi,\pi],\theta_b \in [-\pi/2,\pi/2]$. For a fixed value of $\mathbf{s}$, $\phi(\mathbf{s})$ is constant. Subtracting a deterministic value, does not change the distribution of  $\Delta\phi = \phi(\mathbf{s})-\phi_b$. Finally, $\Delta\phi = [\phi(\mathbf{s})-\phi_b]_{\pi}$ is exploited to ensure wrapping to $[-\pi,\pi]$. Next, using $\Delta \theta = \epsilon_0(\mathbf{s})-\theta_b$, where
$\epsilon_0(\mathbf{s})$ is constant with respect to $\theta_b$, the change
of variable $\theta_b=\epsilon_0(\mathbf{s})-\Delta \theta$ is applied.
Substituting this expression into \eqref{jointPDFangles} yields the desired
distribution. Finally, the range of $\Delta \theta$ is obtained directly from the range of $\theta_b=\epsilon_0(\mathbf{s})-\Delta \theta$, as in Lemma 1.

\section{Proof of Lemma 2}
Conditioned on $\Omega$, $I_s^{\mathrm{UE\, echo}} , I_s^{\mathrm{BS\, echo}}, I_s^{\mathrm{BS\, RF}}$ are conditionally independent terms. Therefore, we can derive $\mathcal{L}_{I_s^{\rm{tot}}}(s|\Omega)$ as the product of conditionally independent Laplace transforms, i.e.,  $\mathcal{L}_{I_s^{\rm{tot}}}(s|\Omega) = \mathcal{L}_{I_s^{\mathrm{UE\, echo}}}(s|\mathbf{p}_0,\Phi_{\mathrm{UE}}^!)  \mathcal{L}_{I_s^\mathrm{BS\, RF}}(s|\mathbf{p}_0,\Phi_{\mathrm{BS}}) \mathcal{L}_{I_s^\mathrm{BS\, echo}}(s|\mathbf{p}_0,\Phi_{\mathrm{BS}})$. First, the proof for deriving $\mathcal{L}_{I_s^{\mathrm{UE\, echo}}}(s|\mathbf{p}_0,\Phi_{\mathrm{UE}}^!)$ is derived as in \eqref{eqLTproof}, shown at the bottom of the next page, where (a) follows from the  moment generating function (MGF) of $h_{\mathrm{UE},i}^s$, (b) follows after averaging over the RCS $\sigma_{\mathrm{UE},i}$ of the UAV-UEs and by setting $P_i^{\mathrm{UE}}(\mathbf{p}_0,\mathbf{p}_i)$ as defined in Lemma 2. This RCS averaging step is justified by the statistical independence between
$\sigma_{\mathrm{BS},j}$ and $\sigma_{\mathrm{UE},i}$, which allows the
corresponding expectations to be factorized. Next, let $u = \sigma_{\mathrm{UE},i}/\overline{\sigma}$ denote the change of variable. Then, $\mathrm{d}\sigma_{\mathrm{UE},i} = \overline{\sigma} \mathrm{d}u$. After substituting $\mathrm{d}\sigma_{\mathrm{UE},i}$ and $\sigma_{\mathrm{UE},i}$ in \eqref{eqLTproof}, $\Psi_m(\cdot)$ yields with argument $x = \frac{s P_i^{\mathrm{UE}}(\mathbf{p}_0,\mathbf{p}_i) \overline{\sigma}}{m}$. Following similar mathematical procedure,  $\mathcal{L}_{I_s^\mathrm{BS\, echo}}(s|\mathbf{p}_0,\Phi_{\mathrm{BS}})$ yields. In a similar vein, $\mathcal{L}_{I_s^\mathrm{BS\, RF}}(s|\mathbf{p}_0,\Phi_{\mathrm{BS}})$ is given by  \eqref{eqLTproof2}, shown at the bottom of the  page, where the $\Phi_{\mathrm{BS}}$-dependent  $\Delta\theta_j^{\mathrm{0 \to BS}}, \Delta\phi_j^{\mathrm{0 \to BS}}$ are involved into $g_{j}^{\mathrm{BS \rightarrow 0}}$. After averaging over the random angular distances  $\Delta\theta_j^{\mathrm{BS \to 0 }}, \Delta\phi_j^{\mathrm{BS \to 0}}$ with the joint PDF $f_{\Delta\phi_{j}^{\mathrm{BS}\to 0},\Delta\theta_{j}^{\mathrm{BS}\to0} \mid\Phi_{\mathrm{BS}}}(\Delta\phi,\Delta\theta)$ given by Lemma 1, $\mathcal{L}_{I_s^\mathrm{BS\, RF}}(s|\mathbf{p}_0,\Phi_{\mathrm{BS}})$  yields. Clearly, $\mathcal{L}_{I_s^\mathrm{BS\, RF}}(s|\mathbf{p}_0,\Phi_{\mathrm{BS}})$ depends on the locations $\mathbf{q}_j$ of $\Phi_{\mathrm{BS}}$. After substituting all the conditional Laplace transforms in $\mathcal{L}_{I_s^{\rm{tot}}}(s|\Omega)$, Lemma 2 yields. 

\begin{figure*}[!b]
\hrulefill
\begin{equation} \label{eqLTproof}
\begin{split}
&\mathcal{L}_{I_s^{\mathrm{UE\, echo}}}(s|\mathbf{p}_0,\Phi_{\mathrm{UE}}^!)=\mathbb{E}_{I_s^{\mathrm{UE\, echo}}}[e^{-s I_s^{\mathrm{UE\, echo}}}] = \mathbb{E}_{h_{\mathrm{UE},i}^s,\sigma_{\mathrm{UE},i}}\left[\exp\left(-s  \sum\limits_{\mathbf{p}_i \in \Phi_{\mathrm{UE}}^!}  (g_{i}^{\mathrm{0 \rightarrow UE}})^2 h_{\mathrm{UE},i}^s \sigma_{\mathrm{UE},i}  l(\|\mathbf{r}_i^{\mathrm{UE}}\|)^2\frac{\lambda^2}{(4 \pi)^3}\right)\right] \\
&\myeqa \mathbb{E}_{\sigma_{\mathrm{UE},i}}\left[\prod_{\mathbf{p}_i \in \Phi_{\mathrm{UE}}^!}\left[  \left(1+\frac{s\,  (g_{i}^{\mathrm{0 \to UE}})^2  \sigma_{\mathrm{UE},i}   \frac{\lambda^2}{(4 \pi)^3}}{m \, l(\|\mathbf{r}_i^{\mathrm{UE}}\|)^{-2}} \right)^{-m} \right] \right] \myeqb \prod_{\mathbf{p}_i \in \Phi_{\mathrm{UE}}^!}\left[ \int_{0}^{\infty}\left(1+\frac{s\,   P_i^{\mathrm{UE}}(\mathbf{p}_0,\mathbf{p}_i) }{m \, \sigma_{\mathrm{UE},i}^{-1}} \right)^{-m} f_{\sigma}(\sigma_{\mathrm{UE},i}) {\rm{d}}  \sigma_{\mathrm{UE},i}  \right]  , \\
\end{split}
\end{equation}
\begin{equation} \label{eqLTproof2}
   \mathcal{L}_{I_s^\mathrm{BS\, RF}}(s|\mathbf{p}_0,\Phi_{\mathrm{BS}})= \mathbb{E}_{\Delta\theta_j^{\mathrm{0 \to BS}},\Delta\phi_j^{\mathrm{0 \to BS}}
}\left[\prod_{\mathbf{q}_j \in \Phi_{\mathrm{BS}}}\left[  \left(1+\frac{s\, g_{j}^{\mathrm{0 \rightarrow BS}} g_{j}^{\mathrm{BS \rightarrow 0}}  K_f l(\|\mathbf{r}_j^{\mathrm{BS}}\|)}{m} \right)^{-m} \right] \right],
\end{equation}
\hrulefill
\begin{equation} \label{proofTh1}
\begin{split}
&\mathcal{P}_{cs}(\theta_c, \theta_s) = \mathbb{E}[\mathds{1}({\rm{SIR}}_{c} \geq \theta_c) \times \mathds{1}({\rm{SIR}}_{s} \geq \theta_s)] \\
&\myeqa  \mathbb{E}_{\mathbf{p}_0,\Phi_{\mathrm{UE}}^!,\Phi_{\mathrm{BS}}}[\mathds{1}({\rm{SIR}}_{c} \geq \theta_c |\mathbf{p}_0,\Phi_{\mathrm{UE}}^!) \times \mathds{1}({\rm{SIR}}_{s} \geq \theta_s |\mathbf{p}_0,\Phi_{\mathrm{UE}}^!,\Phi_{\mathrm{BS}}] \\
& \myeqb \mathbb{E}_{\mathbf{p}_0,\Phi_{\mathrm{UE}}^!,\Phi_{\mathrm{BS}}}[\mathbb{P}({\rm{SIR}}_{c} \geq \theta_c |\mathbf{p}_0,\Phi_{\mathrm{UE}}^!) \times \mathbb{P}({\rm{SIR}}_{s} \geq \theta_s |\mathbf{p}_0,\Phi_{\mathrm{UE}}^!,\Phi_{\mathrm{BS}})], \\
& \myeqc \mathbb{E}_{\mathbf{p}_0} \left[ (f_{\mathbf{p}_i}(\mathbf{p}_i))^{{N_{\mathrm{UE}}-1}} (f_{\mathbf{q}_j}(\mathbf{q}_j))^{N{\mathrm{BS}}}  \prod_{i=1}^{{N_{\mathrm{UE}}-1}} \int_{\mathcal{C}}\mathrm{d} \mathbf{p}_i \prod_{j=1}^{
 {N_{\mathrm{BS}}}} \int_{\mathcal{C}}\mathrm{d} \mathbf{q}_j \mathcal{P}_{s}(\theta_s|\mathbf{p}_0,\Phi_{\mathrm{UE}}^!,\Phi_{\mathrm{BS}})\mathcal{P}_{c}(\theta_c|\mathbf{p}_0,\Phi_{\mathrm{UE}}^!)  \right]
\end{split}
\end{equation}
\end{figure*}

\section{Proof of Theorem 1}
Recalling Definition 1, the ISAC coverage probability is given by \eqref{proofTh1}, shown at the bottom of the page, 
where (a) follows directly from the independence of  $\mathbf{p}_0, \Phi_{\mathrm{UE}}^!$ and $\Phi_{\mathrm{BS}}$. Therefore, conditioned on $\mathbf{p}_0, \Phi_{\mathrm{UE}}^!, \Phi_{\mathrm{BS}}$, the sensing and communication term, are independent random variables. Next, (b) follows after applying the indicator function which is equivalent to  calculating the product of the conditional coverage probabilities. The latter have been proved in Proposition 1 and Proposition 2. Finally, (c) follows after substituting Proposition 1 and Proposition 2 in \eqref{proofTh1} and applying the expectation on \eqref{proofTh1} to jointly decondition over the locations $\mathbf{p}_i$ and $\mathbf{q}_j$ of the interfering UAV-UEs and UAV-BSs, respectively  w.r.t $f_{\mathbf{p}_i}(\mathbf{p}_i)$ and $f_{\mathbf{j}_j}(\mathbf{q}_j)$. Then, Theorem 1 is obtained after deconditioning over the shared location $\mathbf{p}_0$ of the target UAV-UE.

\balance
\bibliographystyle{IEEEtran}
\bibliography{IEEEabrv,references}

\end{document}